\documentclass[fleqn,usenatbib]{mnras}

\usepackage{newtxtext,newtxmath,hyperref}
\usepackage{lastpage}
\usepackage{soul}

\usepackage[T1]{fontenc}

\DeclareRobustCommand{\VAN}[3]{#2}
\let\VANthebibliography\thebibliography
\def\thebibliography{\DeclareRobustCommand{\VAN}[3]{##3}\VANthebibliography}

\usepackage{graphicx}	
\usepackage{amsmath}	
\usepackage{subcaption}
\usepackage{booktabs}
\usepackage[dvipsnames]{xcolor}
\usepackage[normalem]{ulem}
\definecolor{babypink}{rgb}{0.96, 0.76, 0.76}

\newcommand*{\msun}{\ensuremath{\text{M}_\odot}}

\newcommand{\jsubs}{\vec{J}_{\mathrm{Sub}}}
\newcommand{\jhost}{\vec{J}_{\mathrm{Host}}}
\newcommand{\jtot}{\vec{J}_{\mathrm{Tot}}}
\newcommand{\Asubs}{\vec{A}_{\mathrm{Sub}}}

\newcommand{\Ahost}{\vec{A}_{\mathrm{Host}}}

\newcommand{\costheta}{\rm cos(\theta)}

\newcommand*{\mailto}[1]{\href{mailto:#1}{#1}}
\newcommand*{\http}[1]{\href{http://#1}{#1}}
\newcommand*{\https}[1]{\href{https://#1}{#1}}

\defcitealias{fielder2020}{F20}

\begin{document}

\title[Subhalos are Disributed Along Host Major Axis]{Subhalos are Anisotropically Distributed and Aligned with the Smooth Matter Distribution of Their Host Halos}

\author[L.~Mezini et al.]{%
Lorena~Mezini,$^{1}$\thanks{E-mail: \mailto{lom31@pitt.edu}}
Andrew R. Zentner,$^{1,2}$ 
Kuan Wang,$^{3,4}$
Catherine Fielder, $^{5}$
\\
$^{1}$Department of Physics and Astronomy, University of Pittsburgh, Pittsburgh, PA 15260, USA\\
$^{2}$Pittsburgh Particle Physics, Astrophysics, and Cosmology Center (PITT PACC), University of Pittsburgh, Pittsburgh, PA 15260, USA\\
$^{3}$Department of Physics, University of Michigan, Ann Arbor, MI 48109, USA\\
$^{4}$Leinweber Center for Theoretical Physics, University of Michigan, Ann Arbor, MI 48109, USA\\
$^{5}$Steward Observatory, University of Arizona, 933 North Cherry Avenue, Rm. N204, Tucson, AZ 85721-0065, USA
}

\date{Accepted XXX. Received YYY; in original form ZZZ}

\pubyear{2022}

\label{firstpage}
\pagerange{\pageref{firstpage}--\pageref{lastpage}}
\maketitle

\begin{abstract}
We investigate the distributions of subhalos about their hosts in two suites of 
zoom-in N-body simulations of halo growth -- one suite focused on Milky Way-Mass 
halos ($\sim 10^{12} \mathrm{M}_{\odot}$) and another focused 
on cluster-mass halos ($\sim 10^{15} \mathrm{M}_{\odot}$) in the Symphony simulation suite. 
We find, in agreement with previous work on this subject, 
that subhalos are distributed anisotropically about their host halos.
In particular, the positions of subhalos lie preferentially near the major axes of their 
host halos, possibly implying that satellite galaxies will exhibit a similar alignment.
Furthermore, we show that in two-dimensional projection subhalos are more likely to be observed near the halo center (where the central galaxy presumably resides) when the host halo is projected nearly along its major axis.
This projection effect is significant. Within projected radii 
of a few percent of the virial radius of the host halo, the fraction of mass in subhalos is $\sim 175\%$ larger for Milky Way mass halos and as much as $\sim 195\%$ larger for cluster halos when projected along the major axis
as compared to the average from a random projection.
This result has consequences for many applications including the 
interpretation of gravitational lenses.
Finally, we find that the orbital angular momentum vector of subhalos is 
aligned with the angular momentum vector of their host halo, 
indicating that a significant component of a halo's angular momentum may 
be carried in its subhalos.
This has consequences for galaxy formation models 
which use host halo angular momentum as a proxy for galaxy momentum.
\end{abstract}

\begin{keywords}
    galaxies: haloes -- large-scale structure of Universe -- cosmology: dark matter -- galaxies: clusters: general -- gravitational lensing: strong
\end{keywords}

\section{Introduction}

The contemporary galaxy formation paradigm holds that galaxies form within the potential wells 
supplied by halos of dark matter \citep{white1978,blumenthal1984}. 
In the standard $\Lambda$CDM (cold dark matter with a cosmological constant) 
model of the universe, dark matter halos form when perturbations 
in the initial matter field collapse into gravitationally-bound objects. 
These structures grow by the accretion of mass from surrounding dark matter 
and through mergers with other halos; smaller halos that are not entirely 
disrupted upon merging with a larger halo are known as subhalos 
(see e.g., \citealt{kauffmann1993, klypin1999, moore1999, taylor2002, zentner2003, taylor_babul2004, ZentnerBerlind2005, bullock2010, jiang2016, vandenbosch2014}).

$\Lambda$CDM predicts the existence 
of subhalos on all mass scales. 
Some alternative models for dark matter predict differing amounts of substructure.  
Warm dark matter (WDM) has 
become the benchmark example of an alternative to CDM that predicts less 
substructure in general and 
fewer subhalos in particular 
\citep[e.g.,][]{colin+2000,bode+2001,lovell+2014}. 
However, various alternatives to CDM make 
distinct predictions for the structure and 
abundance of dark matter, including 
self-interacting dark matter 
\citep[e.g.,][]{tulin+yu2018}, 
fuzzy dark matter \citep[e.g.,][]{hui2021}, 
and decaying dark matter 
\citep[e.g.,][]{peter_benson2010,wang+2013,wang+2014}. 
Consequently, the observation 
(or non-observation) of subhalos within larger 
host halos can be used as a test of dark matter. 

One tool for detecting dark matter substructure is strong gravitational 
lensing -- a phenomena which takes place when a massive foreground object, 
such as a dark matter halo and the galaxy that it contains, 
deflects the light of a background luminous object, such as a galaxy or quasar. 
For strong lensing to occur, the two-dimensional projected density of the ``lens'' must be greater 
than a critical density. This threshold is often only reached in the area 
near the centers of halos with size defined by the \textit{Einstein radius}.
For substructure lensing to occur, in particular, 
the projected positions of subhalos must be near the scale of the Einstein radius.
For example, lensing by substructure has been noted in multiply-imaged systems 
\citep{Meneghetti2006,Alard2008,Hezaveh2016} 
as well as arc systems \citep{Mao_1998, Metcalf_2001,Brada_2002,Metcalf_2002,Chiba_2002,Dalal_2002, Keeton_2003,Metcalf_2004,Oguri2012}.
If subhalos are distributed anisotropically with respect to their host halos,
it is more likely that subhalos will fall in close proximity 
to the Einstein radius when a halo is viewed at a specific orientation. 
In other words, the probability of lensing by substructure is not uniform, 
but instead is biased by halo shape and halo orientation. 
The goal of this work is to explore the anisotropic distribution of 
subhalos about their hosts halos as a prelude to forthcoming 
studies of strong gravitational lensing. However, the anisotropic 
distributions of subhalos (and, by assumption, of satellite galaxies) 
about their hosts can have consequences for many kinds of observations, 
some of which we discuss briefly below.

A significant body of work predicts that subhalos in CDM simulations of 
structure formation should be distibuted anisotropically about their 
host halos. Host dark matter halos are ellipsoidal in shape \citep{barnes1987,dubinkski_1991,warren_1992,Dubinski_1994,tormen1997,Thomas_1998,Jing2002,springel2004,Hopkins2005}, 
and in particular, halos tend to be prolate 
ellipsoids. \citealt{zentner2005} found preferential alignments between the overall mass 
distribution of Milky Way-mass host halos and the spatial distribution of their subhalos, 
and this was confirmed in galaxy formation simulations shortly thereafter \citep{Libeskind_2007}. 
More recently, \citet{Karp_2023} find an overdensity of satellite galaxies
along the major axis of their central galaxy in the IllustrisTNG simulation. 
It is believed that these phenomena are the result of halos preferentially merging 
along filaments in the large-scale structure \citep{Wang_2005, zentner2005, Libeskind_2007, libeskind_2011, Libeskind_2015, shi2015, Kang_2015, shao_2018, Morinaga_2020}. These filaments trace out 
preferred directions about halos in the large scale structure. 
Several simulations have found that alignments between neighboring dark matter halos tend 
to take place along the direction of the longest axes (the major axes) of the halos. 
\citealt{Faltenbacher_2002} and \citealt{Hopkins2005} detect alignments between the major axes 
of cluster-sized halos out to separations of approximately 100~$\textrm h^{-1}$~Mpc and 30~$ \textrm h^{-1}$~Mpc, respectively. Using the IllustrisTNG simulation, \citet{Rodriguez_2024} 
reports alignments out to 10~$\textrm h^{-1}$~Mpc for both the stellar 
and dark matter components of galaxies.

The prediction that subhalos are disributed anisotropically about their hosts and that they 
are aligned preferentially with the non-spherical dark matter distribution of their hosts 
has some observational support from a variety of analyses of galaxies from surveys 
such as the Sloan Digital Sky Survey (SDSS).
Using a sample of galaxy groups from SDSS DR4, \citet{Brainerd_2005} and \citet{Yang_2006} showed that satellites tend to be aligned with the major axis of the central galaxy in the group. \citet{Azzaro+07} reached qualitatively similar 
conclusions for the SDSS DR4 sample using complementary methods \citep[see also][]{Azzaro+06}. 
\citet{Wang_2018} and \citet{Rodriguez_2022} performed a similar analyses using SDSS DR7 and SDSS DR16 data respectively and found that this alignment exists primarily between red central galaxies and their red satellites. More recently, \citet{samuels_2023} showed that satellite distributions tend to be lopsided with satellites preferentially residing on one side of their host in agreement with \citet{libeskind_2016} and \citet{brainerd2021}. 
Similar results have been observed in cluster lensing analyses. 
Using SDSS \citet{evans_bridle2009}, determined that dark matter halos are
preferentially aligned in the same direction as indicated by the 
spatial distribution of cluster galaxies using cluster lensing. 
\citet{gonzalez+2021} drew similar conclusions from a galaxy lensing analysis 
of a number of surveys.

In contrast to the large-scale structure survey analyses 
mentioned above, satellites of the Milky Way and M31 galaxies have 
been observed to preferentially lie in a plane \textit{perpendicular} 
to their host galaxy major axis \citep{lynden-bell_1976, kunkel_1976, majewski_1994, mateo_1998, grebel_1998, Hartwick_2000,Willman_2004, kroupa_2005,metz_2009, Pawlowski_2012, conn_2013, Ibata_2013}.
This particular alignment of satellites has been dubbed the 
\textit{Holmberg Effect} and was first described in \citet{holmberg1969}. 
Aside from the fact that these Local Group measurements refer to only two central galaxies, 
there are other differences between the Holmberg Effect in the 
Local Group and the SDSS measurements. 
First, and foremost, the central galaxies of the Local Group 
are both disk galaxies, whereas the statistics of the SDSS measurements 
were dominated by elliptical galaxies. 
Second, the Milky Way and M31 form a close pair of galaxies and 
it is uncertain the degree to which alignment effects may be enhanced 
or reduced in close pair systems.

In addition to the biases in the angular distributions of subhalos about their host 
galaxies, subhalos may also be kinematically biased relative to the smooth dark matter component 
in their hosts. These biases may be relevant in studies of galaxy formation. In models of disk galaxy 
formation, the central disk galaxy acquires its angular momentum from its host halo 
\citep{navarro1991, navarro1993, navarro1997, D_Onghia2006, kaufmann2007}. 
For this reason, host halo angular momentum is often used as a proxy for 
galaxy angular momentum in semi-analytic 
galaxy formation models \citep{Somerville2008,Benson2012}.
However, the influence of subhalos on angular momentum must 
not be overlooked in these models. Subhalos that merge from coherent directions, 
such as along filaments, often have angular momenta that are aligned with the 
angular momentum of their host halo \citep{Aubert_2004}. However, it is not 
clear that the angular momentum carried by the subhalos should be included in the 
available reservoir of angular momentum that can be conveyed to the central 
disk galaxy. If it is the case that subhalos carry more angular momentum per 
mass than hosts in general and that subhalo angular momentum displays 
some preferred alignment with the host halo, this could induce a bias in 
models that estimate the amount of halo angular momentum transmitted 
to the baryonic disk. Whether or not the subhalos represent a distinct 
population and whether or not they represent the angular momentum of typical 
host particles should be accounted for in such models.

In this paper, we revisit the question of the degree to which 
the positions and momentum of subhalos within host halos are aligned with the 
mass distribution of their hosts, particularly, whether or not subhalos 
are preferentially aligned with the principle axes and angular momenta of their hosts. As 
will be evident, the primary application we have in mind is strong 
gravitational lensing, though we have attempted to summarize results in 
such a way as to be useful for a variety of other applications. 
This paper is organized as follows. In \autoref{Section:Simulations} 
of this paper we discuss the simulations used, 
\autoref{Section:axis_calcs} introduces the various halo properties that 
we studied, and in \autoref{Section:hosts} and \autoref{Section:subs} we 
go over the alignment results for host halos and 
host-sub halo alignments, respectively. 
Finally we discuss our results and draw conclusions in \autoref{sec:discussion_conclusions}.

\section{Simulations}
\label{Section:Simulations}

In this section, we briefly discuss the simulation data as well as the procedure that we use to construct 
our various data sets.
The two simulations we describe are part of the
Symphony suite of cosmological zoom-in simulations \citep{Nadler_2023}.
We refer the reader to \citet{Nadler_2023} 
for a more detailed description of the simulations. 
The technique that we use for extracting data sets of 
host halos and subhalos is the same as that used in 
both \citet{Mezini_2023} and \citet{fielder2020}. \
We refer the reader to these earlier works for more detailed 
descriptions of this procedure.

\subsection{Milky Way and Cluster Mass Zoom-In Simulations}

We use two sets of zoom-in cosmological simulations from the Symphony suite, each representing a different halo mass range.
The first set of simulations, which were first presented in \citep{Mao2015}, contains high-resolution zoom-in simulations of 45 Milky Way-mass halos with particle mass $4.0\times 10^{5}\msun$ from a c125-2048 parent box run
with L-GADGET (see \citealt{becker2015}). Halo masses in this catalog are $M_{\rm vir} =  10^{12.09 \pm 0.02}\ \mathrm{M}_{\odot}$. The host halos in this catalog have between 83 $\pm$ 18 subhalos at $z=0$, where the quoted range corresponds to the 1$\sigma$ host-to-host scatter in the number of subhalos.
The second set contains zoom-in simulations of 
96 cluster-mass halos with particle mass $1.8\times 10^{8}\msun$ that are within the mass range 
$10^{14.96\pm 0.03}M_{\odot}$ 
first presented in \citep{wu2013}.The host halos in this catalog have 210 $\pm$ 31 subhalos at $z=0$. We exclude one halo from the cluster-mass set due to a potential mislabeling between host halo and most massive subhalo.

\subsection{Halo Identification}

Dark matter halos in these simulations were identified by the 
$\tt{ROCKSTAR}$ (version 0.99.9-RC3+) halo finder \citep{behroozi2013}. 
In summary, $\tt{ROCKSTAR}$ is a 6D phase-space-based finder that 
produces catalogs of all the halos identified as well as tables of all 
the particles associated with each halo. 
A more detailed description of $\tt{ROCKSTAR}$ can be found in 
\citet{behroozi2013}; the source code is publicly 
available\footnote{\url{https://bitbucket.org/gfcstanford/rockstar}}. 
We use the tables of particles which are produced for each halo 
to analyze subhalo positions and angular momenta relative to that 
of their hosts.

Our $\tt{ROCKSTAR}$ based definition of halo components is the same as 
in the previous works of \citet{fielder2020} and \citet{Mezini_2023} and 
translates into the following three groupings of halo particles.

\begin{itemize}
    \item \textit{host only}: These are particles that are associated with the host halo but \textit{not} associated with any subhalos identified by $\tt{ROCKSTAR}$. The \textit{host only} particles are part of the smooth 
    component of the host halo or diffuse structures such 
    as streams that are not identified by $\tt{ROCKSTAR}$.\\
    \item \textit{subhalo only}: These are particles associated with at least one subhalo within the host halo, as identified by $\tt{ROCKSTAR}$. Subhalos within the symphony simulation are defined such that they contain a minimum of 300 particles at z=0.\\
    \item \textit{total halo}: These are particles associated with the host halo, including all particles associated with subhalos within the host. In the vast majority of analyses in the literature, this is the set of particles that is used to characterize a halo; however, both \citet{fielder2020} and \citet{Mezini_2023} have found that the smooth, \textit{host only} component of the host halo is markedly different from the \textit{total halo} particles in a number of important ways.
\end{itemize}

From the preceding definitions, it follows that 
combining the particles in the \textit{host only} and 
\textit{subhalo only} samples recover the set of 
particles in the \textit{total halo} particle set.

\section{Quantifying Halo Shape and Orientation}
\label{Section:axis_calcs}

\subsection{The Major Axis of the Host Halo}

The shapes of dark matter halos are most often characterized by 
the ratios of the lengths of their principle axes. We will 
refer to the principle semi-axes lengths as $a$, $b$, and $c$, 
so that $a > b > c$. The major axis length is thus $a$.
Dark matter halos are most often found to be 
nearly prolate ellipsoids, which have 
$a > b\sim c$, but they can 
also be triaxial ellipsoids where $a > b > c$, and 
$b$ and $c$ are significantly different from one another. 

The lengths of the principle axes can be computed as 
the eigenvalues of the inertia tensor, while their orientations 
are specified by the eigenvectors. The largest eigenvalue 
specifies $a$, while the smallest eigenvalue is $c$. 
The inertia tensor is 

\begin{equation}
I_{ij} = \sum_{n}x_{i,n}x_{j,n},
\end{equation}
where $x_{i,n}$ and $x_{j,n}$ are the coordinates of the $n^{\rm th}$ particle 
and the sum is over all particles identified with a halo.
All particles in our simulations have identical masses, so there is no need 
to include the particle masses in the sum above.

\begin{figure*}
    \centering
    \begin{subfigure}[b]{0.45\textwidth}
    \includegraphics[width=\textwidth]{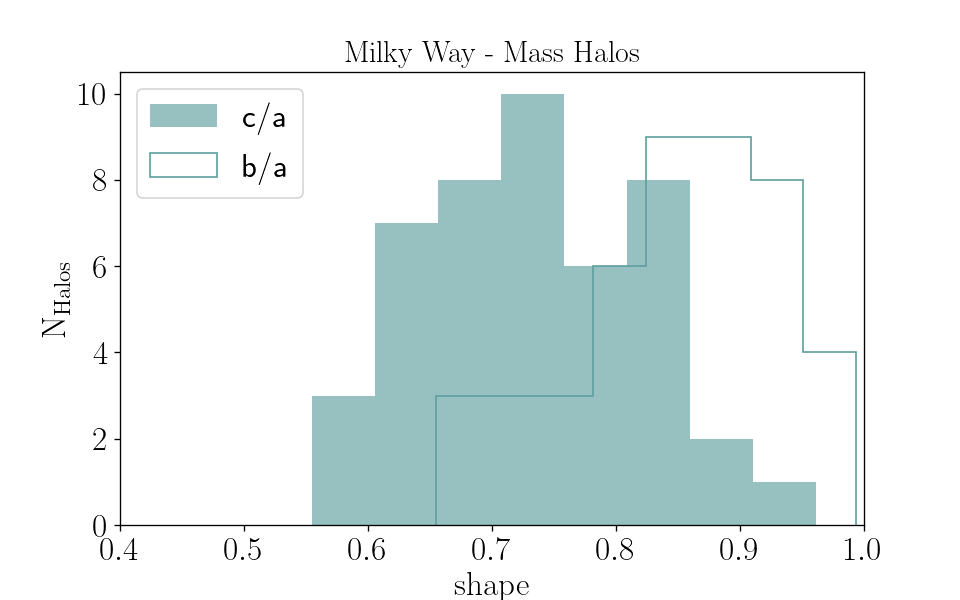}
    \caption{}
    \end{subfigure}
    \begin{subfigure}[b]{0.45\textwidth}
    \includegraphics[width=\textwidth]{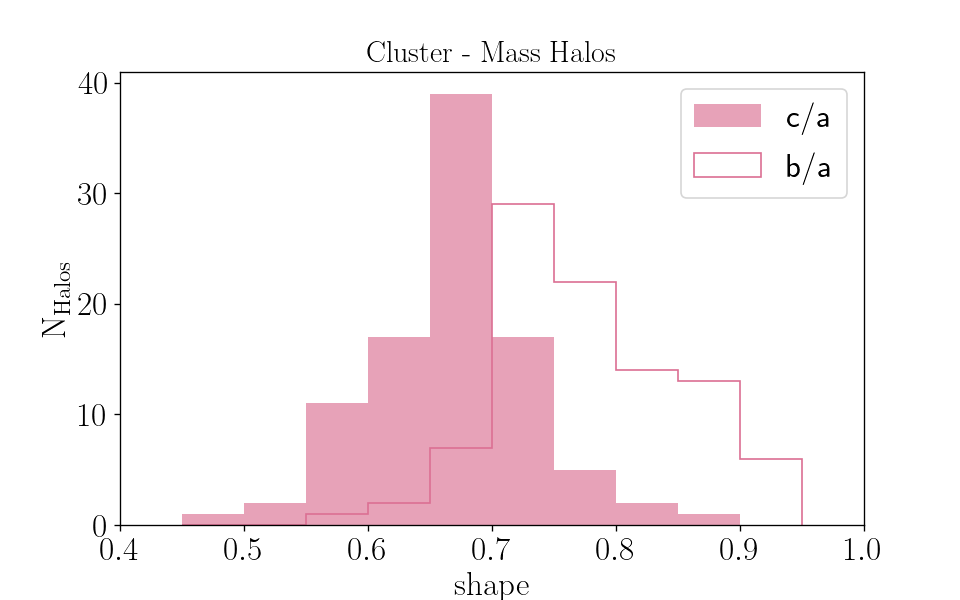}
    \caption{}
    \end{subfigure}
    \caption{Distributions of halo shape as measured by the ratio of the halo minor axis to major axis ($c/a$) and the semi-major to major axis ($b/a$). The filled histograms correspond to c/a and hollow histograms to $b/a$. Milky Way-mass halos are shown in panel ($a$) on the left and cluster-mass halos on right in panel ($b$). Cluster-mass halos are generally less spherical.}
    \label{figure:shape_dist}
\end{figure*}

In figure \autoref{figure:shape_dist}, 
we show the distribution in halo shapes within the Milky Way-mass and Cluster-mass simulations. 
In both figures, the filled histogram corresponds to the ratio between the minor and major axis ($c/a$) and the hollow histogram to the ratio between the semi-major and major axis ($b/a$). 
For both $c/a$ and $b/a$, cluster-mass halos tend to be less spherical compared to Milky Way-mass halos.
Milky Way-mass halos have a median c/a value of 0.74$\pm$0.11 and b/a of 0.85$\pm$0.10.
For cluster-mass halos these values are 0.67$\pm$0.08 and 0.77$\pm$0.1 respectively.
We additionally consider the relationship between halo shape and the proportion of host halo mass locked up in subhalos.
For both mass groups, we calculate the fraction of mass in subhalos for each host and split the host halos into 50th percentile groupings based on the fraction of mass in subhalos. 
We find that the median $c/a$ value is smaller (i.e., halos are more elliptical) for host halos in the group with high subhalo mass fractions; however, the difference is not statistically significant.
For Milky Way-mass halos, the low subhalo mass median shape is 0.78$\pm$0.11 and the high subhalo mass median shape is 0.69$\pm$0.10.
For cluster-mass halos, these values are 0.69$\pm$0.07 and 0.66$\pm$0.08.

We find the directions of the principle axes of the halo by calculating 
the inertia tensor and its eigenvectors directly from the particle data. 
We find the inertia tensor for the \textit{subhalo only} component of 
each halo by subtracting the inertia tensor of the \textit{host only} 
component from that of the \textit{total halo} set. We then diagonalize this inertia tensor to get the subhalo component principle axes.

\subsection{Halo and Subhalo Angular Momentum}
\label{subsection:ang_mom}
Similar to the major axis, angular momentum is also calculated 
directly from particle data. Per particle, we calculate the cross product 
of $\vec{r}_n$, the position of the $n^{\rm th}$ 
particle with respect to the host halo’s center, 
and $\vec{v}_n$, the velocity of $n^{\rm th}$ 
particle, also with respect to the host halo center. 
Summing over the angular momenta of all particles gives the total halo angular momentum, 

\begin{equation}
    \vec{J} = \sum_{n}m_{n}(\vec{r}_{n} \times \vec{v}_{n}), 
    \label{eq:j}
\end{equation}
where $m_{n}$ is the mass of the $n^{\rm th}$ particle. The total angular 
momentum, $\jtot$, includes the angular momentum associated with both host 
halo and subhalo particles and is computed by summing over the 
\textit{total halo} particle set. 
The host angular momentum, $\jhost$, is calculated using \autoref{eq:j} 
by summing over the \textit{host only} particle set, 
while the angular momentum carried by the subhalos, $\jsubs$, 
is defined to be the difference $\jsubs$ = $\jtot$ - $\jhost$.

\subsection{The Subhalo Mass Distribution}
\label{Section:sub_mass_calcs}\

\begin{figure}
    \centering
    \includegraphics[width=1.\linewidth]{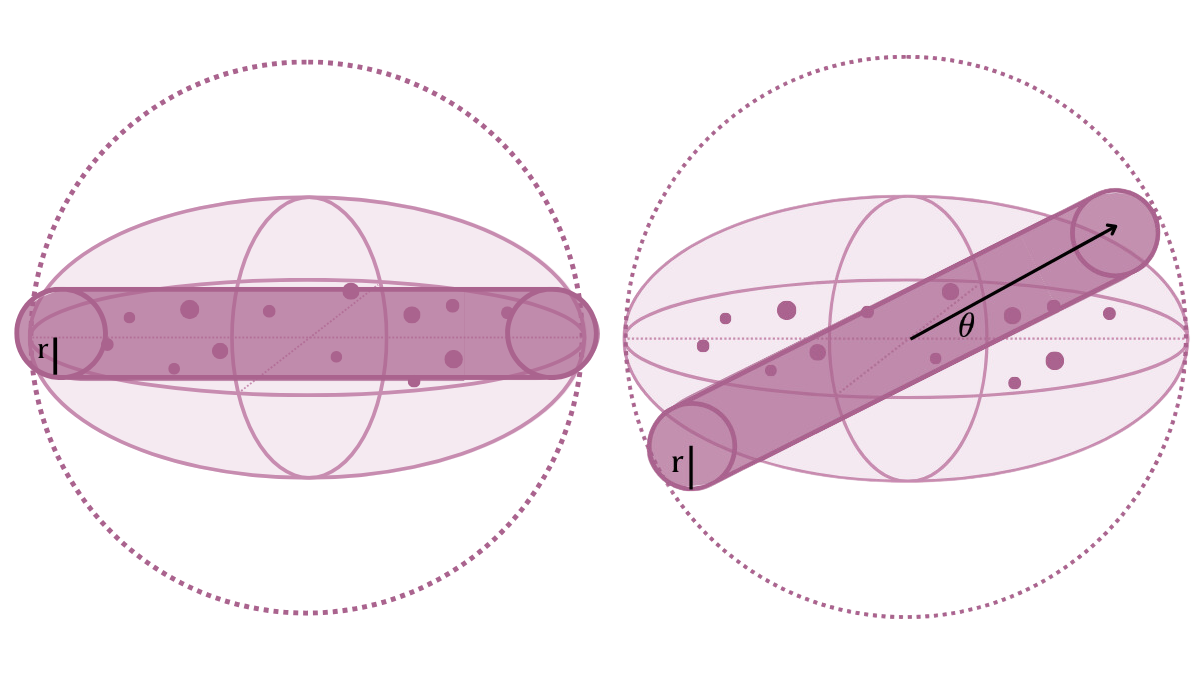}
    \caption{
    Diagram showing how cylinders are defined to calculate projected mass as described in \autoref{Section:sub_mass_calcs}. The dotted pink circle represents the region enclosed within the virial radius of the halo. The light pink ellipsoid represents the shape of the host halo even though all particles within the virial radius are used, and the filled in dark pink circles represent subhalos. The left image shows a cylinder of radius $r$ oriented near the major axis. The cylinder is then rotated to some random orientation as shown in the right image. When comparing the left and right images, we can see how the number of subhalos within cylinders of the same size can vary depending on orientation.
    }
    \label{fig:cylinders}
\end{figure}

We are interested in the way in which the substructure of a 
halo is distributed about its host.
The distribution of subhalos around their hosts in two-dimensional projection and whether this distribution varies based on the projection angle are of special interest due to their relevance to gravitational lensing.
Two-dimensional projections are an interesting special case because the projected 2D density is the relevant quantity in lensing which results in the angular deflections that distort the source.
A number of strong lensing analysis programs count among their goals to 
infer the properties of subhalos (and other line-of-sight halos) in projection \citep[e.g.,][]{Oguri_2010,mckean2015,Ivezi_2019}.
To this end, we calculate the projected mass in subhalos near the 
halo center from different viewing angles, including projections along 
the major axis, for which the projected mass density is highest. 
We recommend referring to \autoref{fig:cylinders} while reading the steps of 
this calculation. Our procedure is as follows.

\begin{enumerate}
    \item Define a set of axes along which to project host halo and subhalo mass. The axes are centered on the host halo and spaced isotropically over the surface of a sphere.
    
    \item Define a cylinder, with a central axis that points along the halo major axis and intersects the halo center such that the circular face of the cylinder has a radius $r < r_{\rm{VIR}}$ and the length of the cylinder extends to the virial radius of the halo.
    \item Measure the total mass of subhalo particles that fall within the 
    cylinder and divide it by the total halo mass to calculate the subhalo mass 
    fraction within the cylinder.
    \item Rotate the cylinder about the halo center so that it points along the subsequent axis in the predefined set of projection axes.
    \item Calculate the angular separation, $\theta$, between the major axis and the new axis of projection.
    \item Repeat the mass measurement from step (iii).
    \item Repeat the process of rotating and measuring the subhalo mass [steps (iii)-(vi)] within each rotated cylinder for all the axes in the predefined set of axis directions from step (i).
\end{enumerate}

This procedure is repeated for each halo of the Milky Way-mass 
and cluster-mass halos separately for cylinders with 
logorithmically-spaced radii from 0.001 to 1 times their host virial radius.
By definition, a radius of $r=0$ encompasses 
none of the halo mass, whereas a cylinder of radius $r=r_{\rm{VIR}}$
encompasses the entire halo.
Once completed, for both cluster-mass and Milky Way-mass halos, 
we will have a distribution of subhalo mass fractions within a cylinder 
of radius $r$ as a function of the angle $\theta$ between the cylinder's 
axis and the halo's major axis.

\section{Results I: Host Halo Alignments}
\label{Section:hosts}

We begin our discussion of halo alignments with a study of 
the properties of particles that are associated with the host halo.
It has been found that the orientation of the halo angular momentum vector 
with respect to the principle axes of the halo is not isotropic. 
Rather, the angular momentum vector is correlated 
with the minor axis of the halo \citep[e.g.,][]{warren_1992,tormen1997}, 
as expected from tidal torque theory \citep[e.g.,][]{ttt1,ttt2}. 
The halo major and minor axes 
are orthogonal to one another (by definition). Therefore, 
this suggests that the angular momentum 
of a halo should be oriented in 
a manner that is preferentially orthogonal 
to its major axis.

\begin{figure*}
    \centering
    \begin{subfigure}[b]{0.4\textwidth}
    \includegraphics[width=\textwidth]{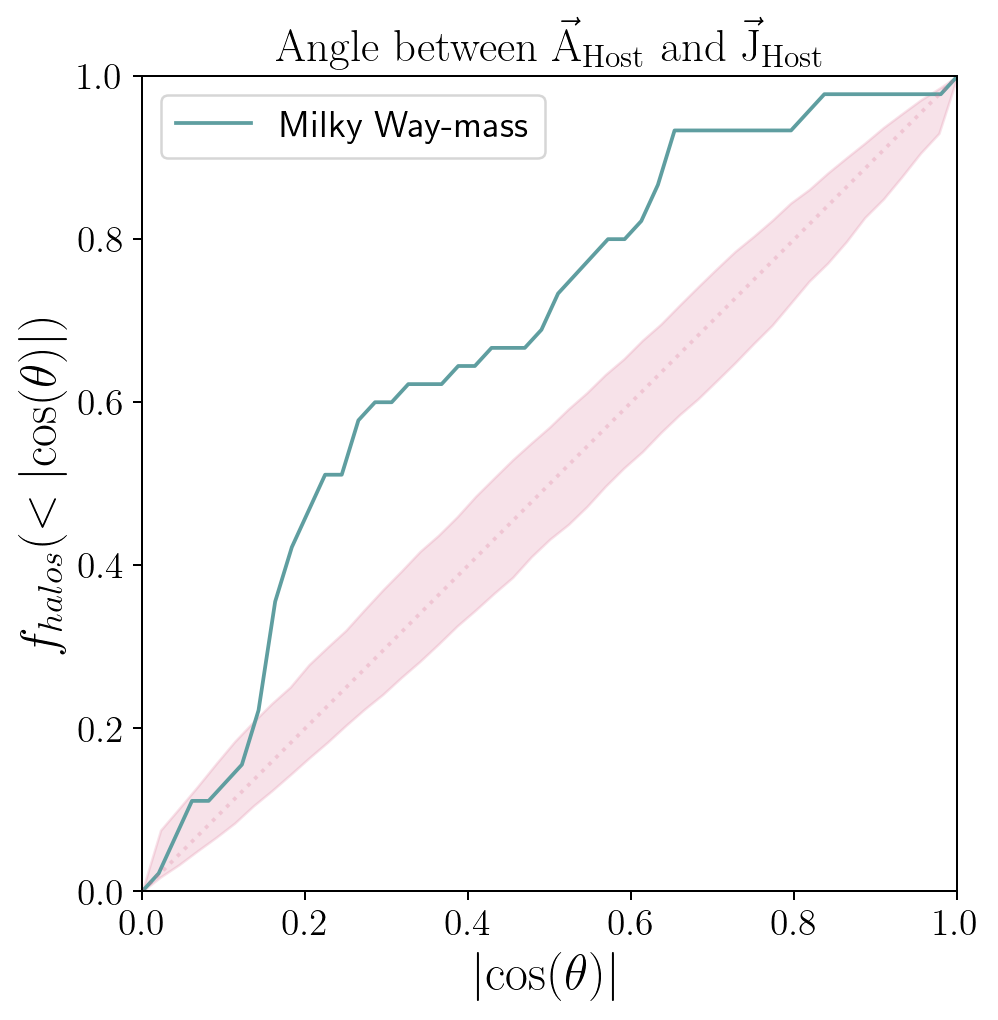}
    \caption{}
    \end{subfigure}
    \begin{subfigure}[b]{0.4\textwidth}
    \includegraphics[width=\textwidth]{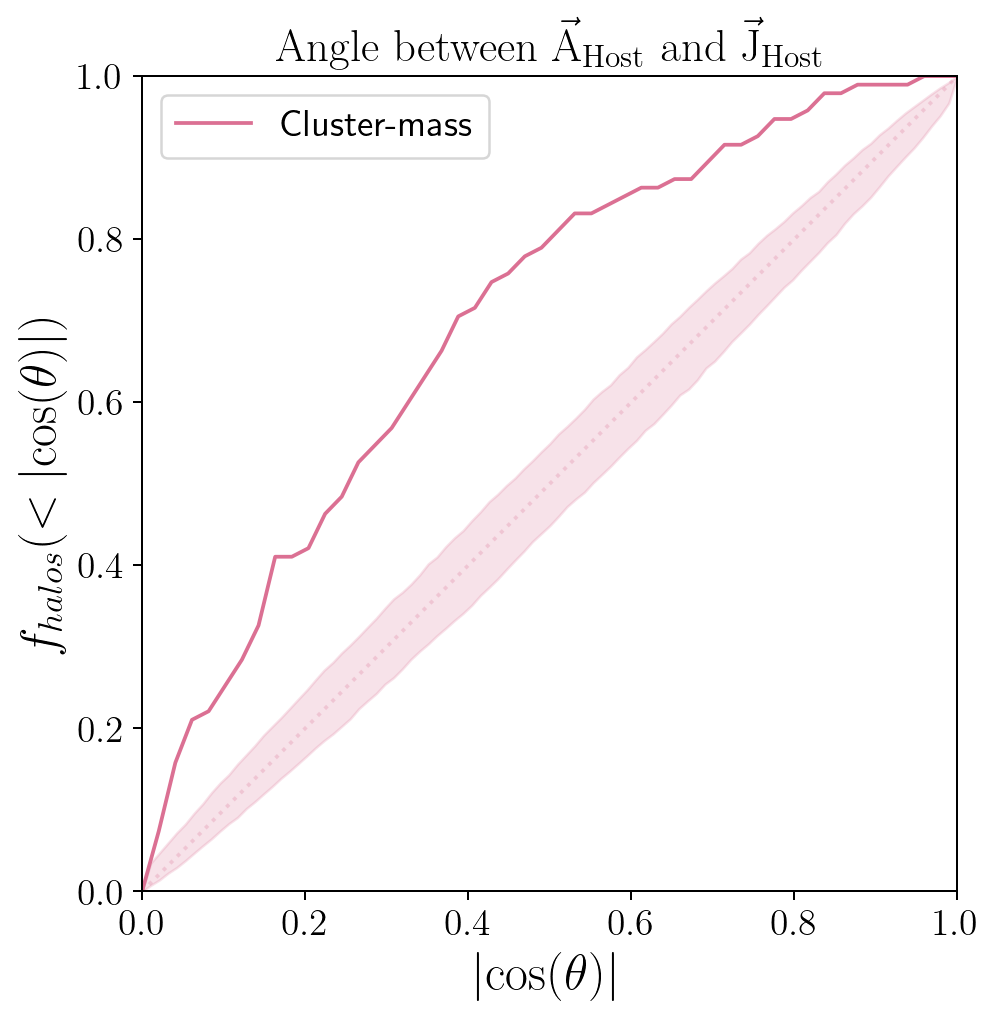}
    \caption{}
    \end{subfigure}

    \caption{The cumulative fraction of halos with angle less than $\costheta$ between the host halo angular momentum vector, $\jhost$ and its major axis $\Ahost$ as a function of $\costheta$. The solid line is the result from the simulation data, the dashed line and the filled area are the CDF for an isotropic distribution and the area within the 14th and 86th percentiles. The percentiles are computed from random samples that are the same size as the simulation samples. Results for Milky Way-mass halos are shown in (a) and cluster-mass in (b).}
    
    \label{figure:cumul_angle_ahost_jhost}
\end{figure*}

In \autoref{figure:cumul_angle_ahost_jhost} we show the cumulative probability distribution of 
halos as a function of the cosine of the angular separation, 
$\vert \cos(\theta) \vert$, where $\theta$ is the angle between 
the host halo angular momentum vector, $\jhost$, 
and the host major axis. The angle $\theta$ plays the 
role of a zenith angle with zero along the angular momentum 
vector, so a model in which angular momenta and major axes are 
uncorrelated would yield a uniform distribution of $\vert \cos(\theta) \vert$. 
Panel (a) corresponds to Milky Way-mass halos and (b) to cluster-mass. 
In each panel, the solid lines represent the cumulative distribution 
function of alignments from our simulation data sets. 
The straight dotted lines represent the distribution expected for 
a random, isotropic relation between angular momentum and major axis. 
The filled areas represent errors relative to an isotropic underlying 
true distribution. These errors were estimated by building 
1000 mock samples of alignments 
from an isotropic underlying distribution. Each mock sample 
contained the same number of data points as the simulation data 
(45 for Milky Way mass and 95 for cluster mass). The errors are 
then the $\rm {16^{th}}$ and $\rm {84^{th}}$ percentiles of these 
mock samples.

As discussed above, we expect that the orientation of $\jhost$ with respect 
to host halo major axes is not isotropic and will be preferentially aligned with 
the minor axes, and perpendicular to the major axes.
Our data is suggestive of an orthogonal alignment between the major axis and angular momentum vector, $\jhost$, as expected. In both panels of \autoref{figure:cumul_angle_ahost_jhost}, 
we see an excess of halos for which the angle between the 
angular momentum vector and the major axis, $\vert \cos(\theta) \vert$, is small, which corresponds to a perpendicular alignment. 
The Kolmogorov-Smirnov (KS) probability that these angular distributions are selected 
from an isotropic distribution is $\sim$ 4 $\times$ 10$^{-\rm 5}$ for Milky Way-mass halos and $\sim$ 3 $\times$ 10$^{-\rm 5}$ for cluster-mass halos, indicating that an isotropic, uncorrelated distribution of halo angular 
momenta and principle axes is excluded by the data. This alignment has previously been detected in works such as \citet{Shaw_2006,Knebe_2008,Deason_2011,Hoffmann_2014,Kiessling_2015}.

We can also quantify the alignment between host halo angular momentum vector and host major axis with the median angular separation between the two. For Milky Way-mass halos, the median angular separation is $\theta$ = $77 \pm 3.0^{\circ} (17^{\circ})$ or $\vert \cos(\theta) \vert$ = 0.22 $\pm$ 0.05 (0.24). The first set of error values are the errors on the medians while the values in the parentheses correspond to the dispersions in the distribution of angles. For Cluster-mass halos, the median angular separation is $\theta$ = $75\pm 2^{\circ} (16^{\circ})$ or cos($\theta$) = 0.25 $\pm$ 0.04 (0.24). Errors are calculated by bootstrapping. For each set of angles, we sample the angles with replacement to create a mock sample of the same size and calculate its median. To get the uncertainty we repeat this 1000 times and calculate the standard deviation of the set of 1000 medians. For an isotropic distribution of angles, we would expect a median of 60$^{\circ}$, however, we are measuring angles that are approximately 15$^{\circ}$ greater. Angles significantly greater than 60$^{\circ}$ are indicative of perpendicular alignment which is consistent with the discussion in the previous paragraph that a halo's angular momentum vector is preferentially perpendicular to its major axis.

\section{Results II: Spatial Distribution of Subhalos}
\label{Section:subs}

Next, we study the spatial distribution of subhalos about their hosts in two ways. 
First, we measure the degree to which the spatial positions of subhalos are 
aligned with the major axis of their host halo and whether or not subhalos 
reside in a planar structure roughly orthogonal to the host angular momentum. 
Second, we measure the way in which the 2d projection of the 
subhalo population belonging to a host changes when the halo is 
viewed from different directions. We quantify this projection effect primarily 
in terms of the projected mass fraction in subhalos.

\subsection{Principle Axis and Angular Momentum Alignments of Subhalos and Their Hosts}
\label{sub_axis_alignment}

\begin{figure*}
    \centering
        \begin{subfigure}[b]{0.4\textwidth}
            \centering
            \includegraphics[width=\linewidth]{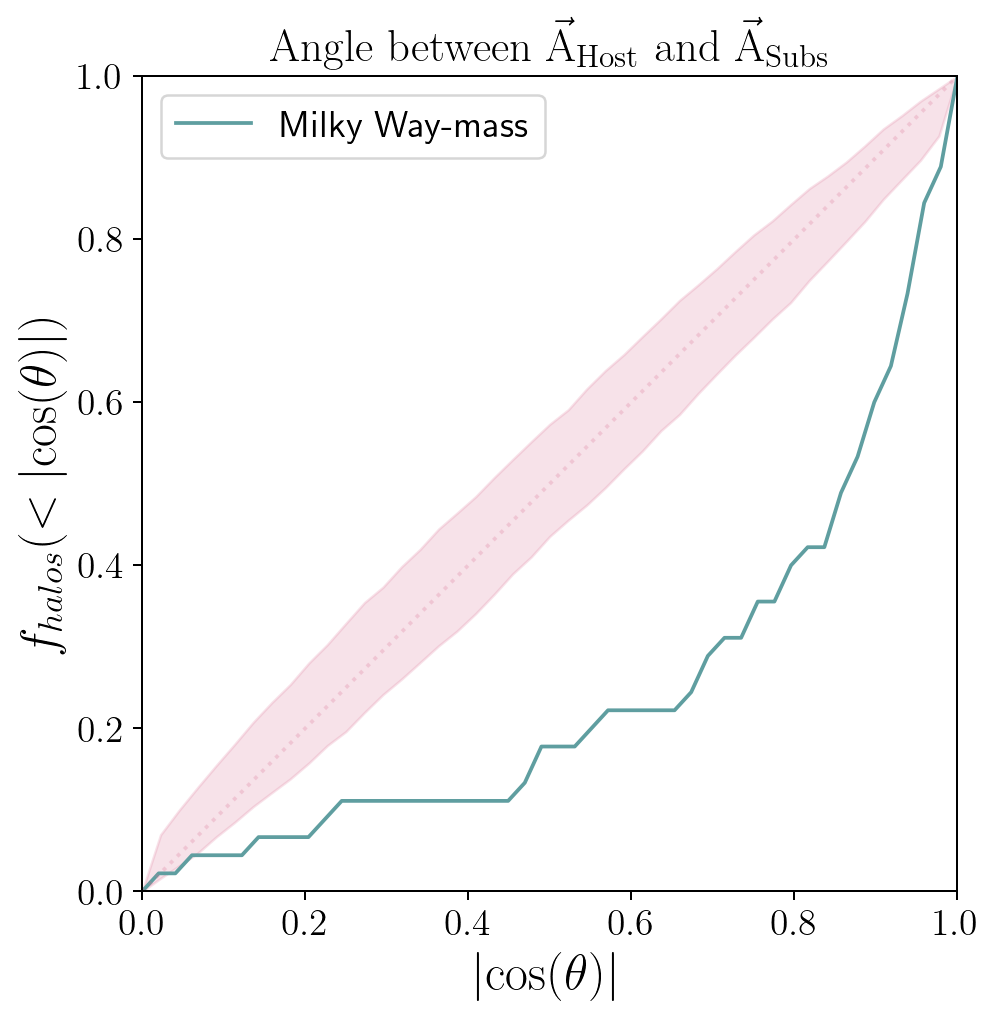}
            \caption{}
            \label{figure:mwm_cumul_asubs_ahost}
        \end{subfigure}
        \begin{subfigure}[b]{0.4\textwidth}
            \centering
            \includegraphics[width=\linewidth]{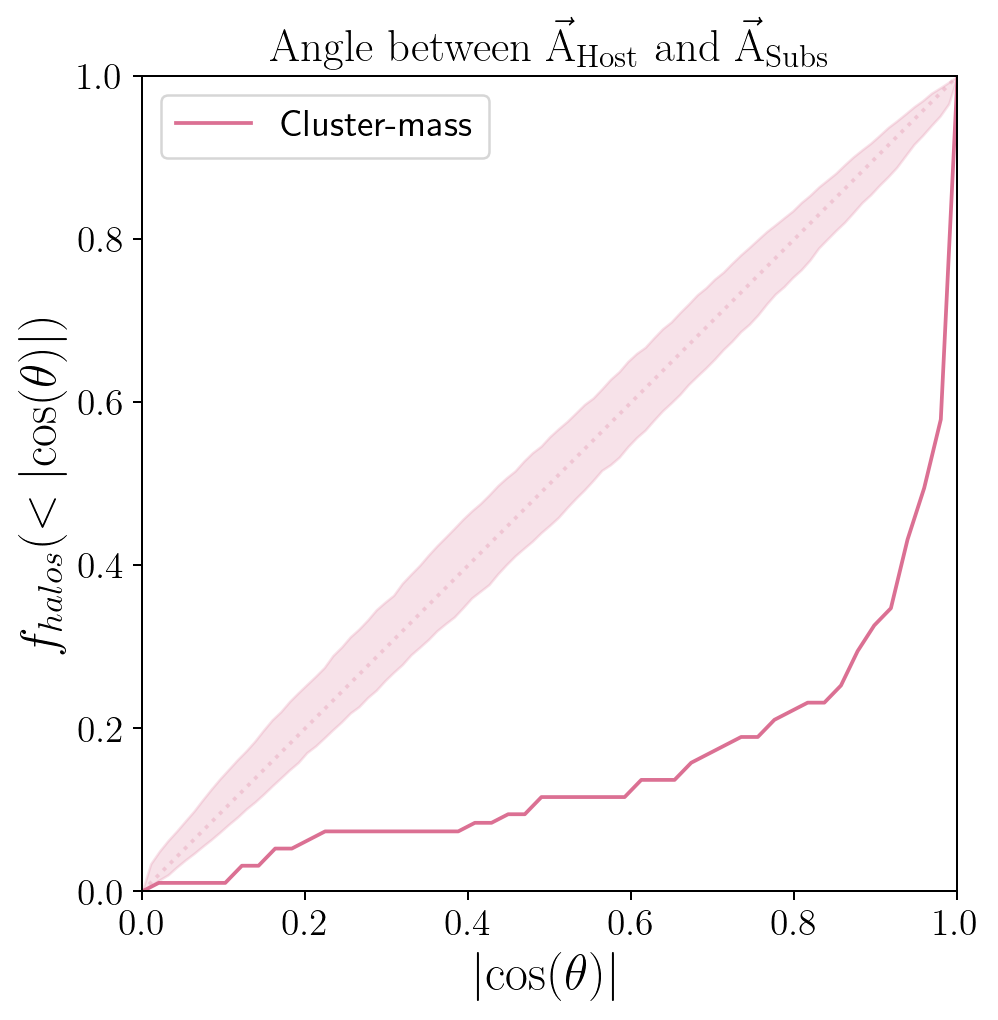}
            \caption{}
            \label{figure:rhap_cumul_asubs_ahost}
        \end{subfigure}
    \caption{The cumulative fraction of halos with angle <$|\rm cos(\theta)|$ between the host halo major axis $\Ahost$ and that of its subhalos, $\Asubs$, as a function of $|\rm cos(\theta)|$. The solid line is the result from the simulation data, the dashed line and the filled area are the CDF for an isotropic distribution and the area within the 14th and 86th percentiles. Panel \textbf{(a)} is for Milky Way mass halos and \textbf{(b)} is for cluster mass halos.}
    \label{figure:cumul_angle_ahost_asubs}
\end{figure*}

We first discuss alignments between the major axis of the distribution of all 
subhalos belonging to a particular host halo and the major axis of said host halo.
The principal axes of the subhalo component are defined as the eigenvectors of the 
inertia tensor corresponding to subhalo particles only.
We calculated the angular separation between the major axis of the host, $\Ahost$, and that of the subhalos, $\Asubs$. Previous works have shown that this result can vary as a function of the radius at which the calculation is performed \citep{Schneider_2012}. In this work, we chose to use major axis as defined up to the full virial radius of the halo.
In \autoref{figure:cumul_angle_ahost_asubs}, we show the cumulative fraction of halos with angle $< |\rm cos(\theta)|$ between $\Ahost$ and $\Asubs$ as a function of $|\rm cos(\theta)|$.
The left plot (\autoref{figure:mwm_cumul_asubs_ahost}) shows results for Milky Way-mass halos and  the right (\autoref{figure:rhap_cumul_asubs_ahost}) for cluster-mass halos.

In both panels of \autoref{figure:cumul_angle_ahost_asubs} we see that there is a deficit of 
halos at smaller values of $|\rm cos(\theta)|$, indicating that the angular separation between 
the major axis of the host halo matter distribution and the major axis of the subhalo distribution 
is smaller than it would be given an isotropic distribution. This implies, perhaps unsurprisingly, 
that $\Ahost$ and $\Asubs$ are preferentially aligned. In other words, the distribution of 
subhalos is aligned with the distribution of host halo mass. The KS probability that the result 
for Milky Way-mass halos (\autoref{figure:mwm_cumul_asubs_ahost}) is selected from an 
isotropic distribution is $P_{\rm KS}$ is $\sim 3 \times 10^{-\rm 9}$. 
For cluster-mass halos (\autoref{figure:rhap_cumul_asubs_ahost}) this probability is 
$\sim 7 \times 10^{-10}$. 

For Milky Way-mass halos, the median angle between the host halo major axis the subhalo component major axis is $\theta$ = $30\pm 4.0^{\circ} (22^{\circ})$ or $\vert \cos(\theta) \vert$ = 0.86 $\pm$ 0.04 (0.26). For Cluster-mass halos, this median angle is $\theta$ = $16\pm 3^{\circ} (23^{\circ})$ or $\vert \cos(\theta) \vert$ = 0.96 $\pm$ 0.01 (0.24). Errors are calculated with the same bootstrap method as \autoref{Section:hosts}. For an isotropic distribution of angles (spanning from 0$^{\circ}$ to 90$^{\circ}$), we would expect a median of 60$^{\circ}$; however, we measure angular separations that are approximately 30$^{\circ}$- 45$^{\circ}$ smaller. This is consistent with the discussion in the previous paragraph that subhalos preferentially reside closer to the major axis 
of their host halo.

The alignment we find between $\Ahost$ and $\Asubs$ is not unexpected, a 
fact to which we have already alluded. 
Within the model of hierarchical structure growth, 
subhalos merge with host halos in a correlated manner 
along filaments \citep[e.g.,][]{Plionis_2002,Jing2002,Bailin_2005,Faltenbacher2005,zentner2005,Shaw_2006, Libeskind_2007,Welker_2014}. 
In a recent work, \citet{han2023} showed that $\sim 68\%$ of accreted subhalos enter 
their host halo through $\sim 38\%$ of the halo surface area defined at the virial radius. 
This results in alignment between the large-scale filamentary structure and 
that of a host halo \citep[e.g.,][]{Jing2002,Bailin_2005,Faltenbacher2005,zentner2005}. 
In turn, this causes the shape and orientation of halos to be ellipsoids with their 
longest axes pointed along the direction of the filament. 
Consequently, subhalos are preferentially located near this longest axis, 
the major axis, and this is the alignment that we detect.

Comparing \autoref{figure:mwm_cumul_asubs_ahost} with \autoref{figure:rhap_cumul_asubs_ahost}, it is evident that the alignment between $\Ahost$ and $\Asubs$ is more pronounced in cluster-mass halos than in Milky Way-mass halos. This difference may be attributed to the later formation times and larger subhalo abundances of cluster-mass halos, which generally have experienced significantly more recent mergers \citep{Lacey1994,wechsler2002,ZentnerBerlind2005}. Consequently, this means that the major axes of cluster-mass halos have been more recently influenced by the influx of mergers along neighboring filaments. Collectively, these effects lead to subhalos being positioned more closely to the major axis of cluster-mass halos. Such a mass dependence on the strength of alignments has been proposed before for neighboring halos in works such as \citet{Jing_2002,Li_2013, schneider2012, Xia_2017}. A follow-up analysis would be necessary in order to confirm this interpretation.

In principle, it may be possible for the major axes of host halo and subhalo components to be aligned without an anisotropic distribution of subhalos (though this seems inconsistent with 
our other results, see below). Suppose that the distribution of 
subhalos is characterized by principle axis ratios $(b/a)_{\mathrm{subs}}$ and 
$(c/a)_{\mathrm{subs}}$ (from the {\em subhalo only} samples) 
that are significantly larger than the principle axis 
ratios determined by the overall mass of the host halos (from the {\em host only}) samples. 
For example, if $(c/a)_{\mathrm{subs}} \sim 1 \gg (c/a)$ and $(b/a)_{\mathrm{subs}} \sim 1 \gg (b/a)$, 
then the subhalo mass would be distributed nearly isotropically, so that axis alignments 
would be far less noteworthy. However, this is not the case. We find that 
subhalo mass is distributed approximately as anisotropically as its host halo's mass as quantified by 
axis ratios. The axis ratios for host halos in the Milky Way-mass simulations are 
$(b/a) = 0.86 \pm 0.10$ and $(c/a) = 0.74 \pm 0.11$, while the corresponding ratios for subhalo mass 
are $(b/a)_{\mathrm{subs}} = 0.84 \pm 0.16$ and $(c/a)_{\mathrm{subs}} = 0.66 \pm 0.16$. 
Similarly, for cluster mass halos the axis ratios are 
$(b/a) = 0.78 \pm 0.08$ and $(c/a) = 0.68 \pm 0.07$ for host halos while they 
are $(b/a)_{\mathrm{subs}} = 0.60 \pm 0.2$ and $(c/a)_{\mathrm{subs}} = 0.36 \pm 0.18$ 
for the subhalos. Subhalo mass is distributed at least as 
anisotropically as the host halo mass indicating that alignments among 
principle axes designate meaningful alignments of a strongly anisotropic 
distribution of subhalos. In fact, our results suggest that the distribution of subhalo mass is somewhat less spherical than the overall mass distribution 
of a halo.

\begin{figure*}
    \centering
    \begin{subfigure}{0.49\textwidth}
        \includegraphics[width=\linewidth]{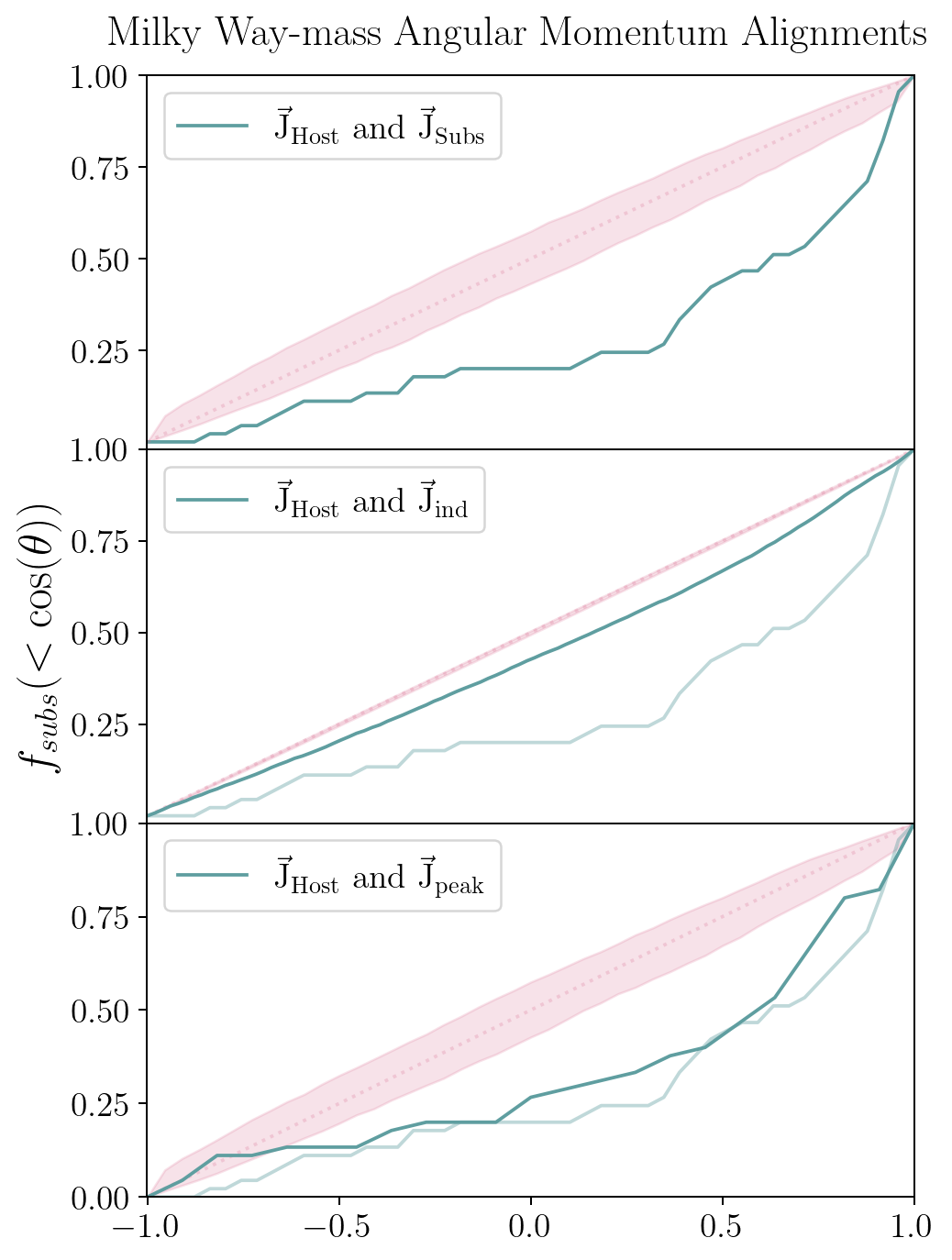}
        \caption{}
        \label{figure:mwm_cumul_angle_jhost_jsubs}
    \end{subfigure}
    \begin{subfigure}{0.49\textwidth}
        \includegraphics[width=\linewidth]{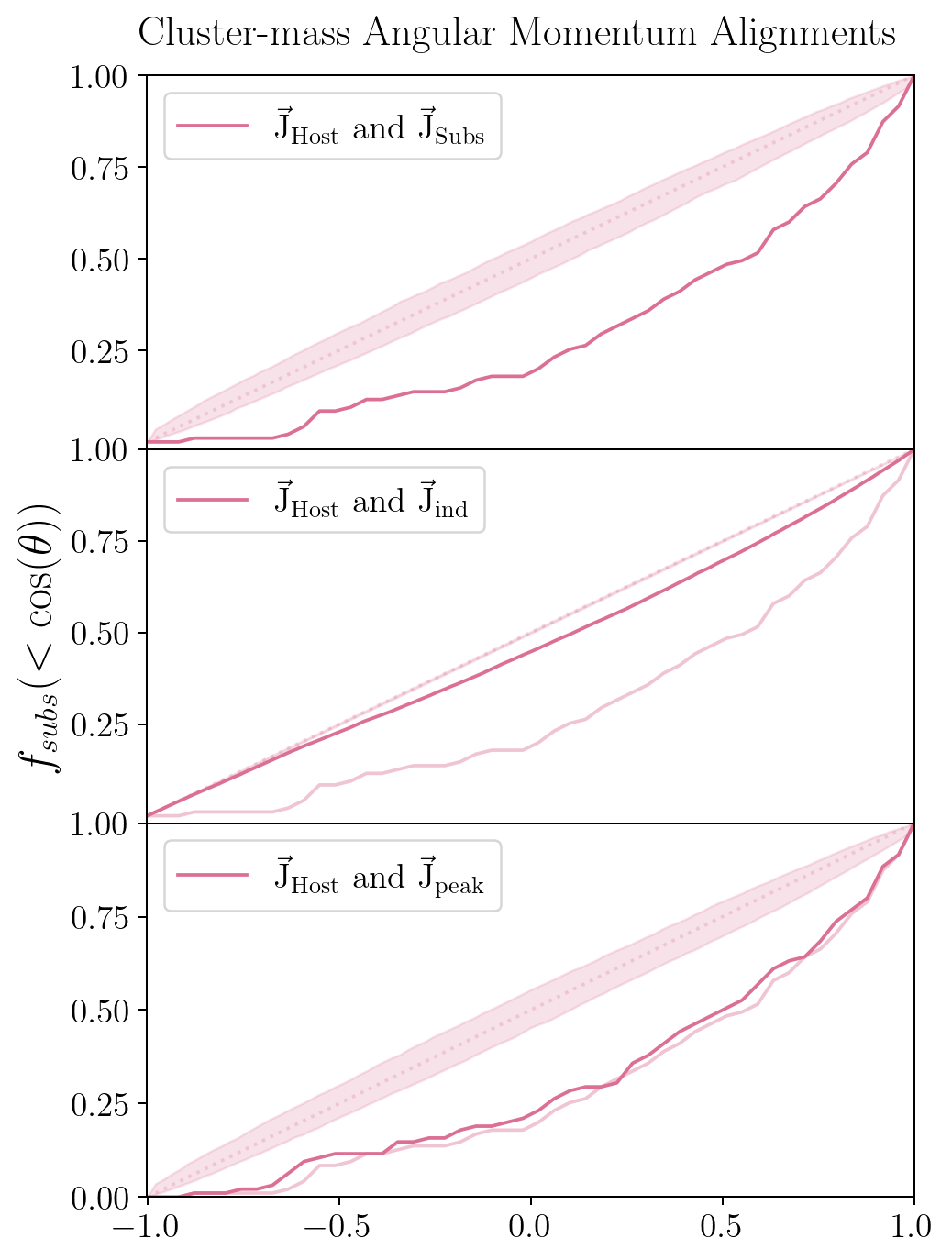}
        \caption{}
        \label{figure:rhap_cumul_angle_jhost_jsubs}
    \end{subfigure}
    \caption{The same as \autoref{figure:cumul_angle_ahost_asubs} except for the angles between the host and subhalo angular momentum vectors, $\jhost$ and $\jsubs$, respectively. In the bottom 4 figures, we use alternative measurements of subhalo angular momentum. For comparison with the original method of measuring subhalo angular momentum, we also plot the lines from the top plots in a lighter shade along side the alternative method result. In the middle row, subhalo angular momenta is defined by the position and velocity of its density peak. Like in the top row figures, we sum over all subhalo angular momenta to define $\vec{J}_{\mathrm{peak}}$. In the bottom row, $\vec{J}_{\mathrm{ind}}$ corresponds to the angular momentum of a single subhalo found by summing over all its particles rather than the sum of particles over all subhalos belonging to a host.}
    \label{figure:cumul_angle_jhost_jsubs}
\end{figure*}

In \autoref{figure:cumul_angle_jhost_jsubs}, we compare the angular momentum vectors of host halos, $\jhost$, to that of their subhalos, $\jsubs$. $\jhost$ and $\jsubs$ are both defined in \autoref{subsection:ang_mom}.
Notice that, unlike orientations relative to the major axis, the x-axis in \autoref{figure:cumul_angle_jhost_jsubs} 
is $\rm cos(\theta)$ rather than $|\rm 
cos(\theta)|$ to display cases in which subhalo angular momenta are anti-parallel 
to that of their host. 
The left panels correspond to Milky Way-mass halos and the right to cluster-mass halos. 
In the top row, figures $a$ and $b$, we see a similar trend as in \autoref{figure:cumul_angle_ahost_asubs} in which there is a 
deficit of halos at smaller values of $\rm cos(\theta)$, indicating that there is alignment 
between $\jhost$ and $\jsubs$. 
The KS probability that the distribution for Milky Way-mass halos in the top panel of \autoref{figure:mwm_cumul_angle_jhost_jsubs} is selected from an isotropic distribution is $P_{\rm KS}$ $\sim 1 \times 10^{-\rm 9}$.
For cluster-mass halos in the top panel of \autoref{figure:rhap_cumul_angle_jhost_jsubs}, $P_{\rm KS}$ is $\sim 4 \times 10^{-\rm 5}$.
As we discussed in the preceding paragraph, 
the tendency for subhalo and host halo angular momenta to be aligned 
makes sense within the context of hierarchical 
growth -- subhalos merging in from coherent directions along 
filaments will preferentially orbit near a plane containing the major axis. 
The fact that we detect alignment between major axes strengthens this argument.

The other two rows of panels in \autoref{figure:cumul_angle_jhost_jsubs} 
show our results for alternative methods of measuring the subhalo angular momentum.
In the first case, we define the angular momentum of each subhalo using the position and velocity of its 
center as specified in the halo catalog and sum over that angular momentum for all subhalos within a host 
to find the total angular momenta of the subhalo component of the halo.
We will refer to this measure of angular momentum as $\vec{J}_{\mathrm{peak}}$. 
This method is distinct from the previous method in which the subhalo angular momenta 
were computed by summing over the particle content of each subhalo. 
In panels $c$ and $d$ we show the cumulative distribution of angles between $\vec{J}_{\mathrm{peak}}$ and $\jhost$. The distributions in $c$ and $d$ are, unsurprisingly, both similar to those of $a$ and $b$. 
This suggests that our alignment results are insensitive to the details of the method 
used to determine subhalo angular momenta. 
In the second alternative case, we display the alignments of angular momenta for \textit{individual} 
subhalos, rather than for populations of subhalos associated with individual hosts. 
In particular, we compute the angle between the angular momentum of 
each individual subhalo and the angular momentum vector of its host. 
To distinguish this measurement, we refer to the angular momentum of a single subhalo as 
$\vec{J}_{\mathrm{ind}}$. We sum over the
angular momenta of all particles in any individual subhalo to determine $\vec{J}_{\mathrm{ind}}$.
The result of this exercise is shown in the bottom panels, $e$ and $f$.
The resulting distribution of alignments differs from those in panels $a$, $b$, $c$, and $d$, 
because in this case the individual subhalos corresponding to a single host can fall 
into distinct bins of zenith angle, whereas in the previous calculation each host had a 
single zenith angle corresponding to the alignment of the host angular momentum with 
the net angular momentum of the population of subhalos. This has the effect of spreading 
the contribution from the subhalo population associated with any particular host galaxy 
among a range of zenith angle values. Furthermore, the alignment distribution computed in 
this way is significantly less noisy because the distribution is assembled from many more 
data points. Calculating angles for individual halos allows us to decipher whether or not 
the alignment in the top two figures, $a$ and $b$, is due to a large fraction of subhalos 
having angular momenta aligned with that of their hosts, or, subhalo angular momenta that are 
spread out but average to a value close to that of its host. The anisotropy is 
weaker for $\vec{J}_{\mathrm{ind}}$, implying that those subhalos with the 
largest angular momenta are more well-aligned with the host angular momentum.

Finally, we quantify the alignment between angular momenta with the median angular separation between host and subhalo angular momentum vectors. For Milky Way-mass halos, the median angular separation is $\theta$ = $50\pm 9^{\circ} (38.0^{\circ})$ or cos($\theta$) = 0.63 $\pm$ 0.12 (0.55). For Cluster-mass halos, it is $\theta$ = $56\pm 5.5^{\circ} (34^{\circ})$ or cos($\theta$) = 0.56 $\pm$ 0.08 (0.49). Errors are calculated by bootstrapping using the same method as in \autoref{Section:hosts}. For an isotropic distribution of angles (spanning from 0$^{\circ}$ to 180$^{\circ}$), we would expect a median of 90$^{\circ}$, with 0$^{\circ}$ indicative of perfect alignment between the angular momentum vectors, and 180$^{\circ}$ indicative of $\jhost$ and $\jsubs$ pointing in complete opposite directions. Our findings show that the median angle of separation between angular momentum of the host and subhalos is significantly smaller than 90$^{\circ}$, indicative of strong alignment between the two, and is consistent with the discussion in the previous paragraph that subhalos should orbit in the same plane as the host halo.

Although we anticipate that the alignments of $\Asubs$ and $\Ahost$, and, $\jsubs$ and $\jhost$ are a result of mergers along coherent directions, such as filaments, we 
should expect significant scatter in these alignments.
Cluster-sized halos are comparable in size to the filament in which they reside and may also be located at a node connecting multiple filaments.
Galaxy-sized halos at present are often smaller than the size of the filaments in which they live but may have accreted their surviving subhalos at times when filaments were thinner \citep{Vera_Ciro_2011}.
The distribution of subhalos that merge from other directions or from presently-subdominant 
filamentary structure introduces considerable scatter into the anisotropic distributions 
we have found. Furthermore, simulations have shown that spin alignment with 
filaments has a mass dependence 
\citep{Bailin_2005,Arag_n_Calvo_2007,Paz2008,zhang2009,Codis2012,libeskind_2011,Libeskind_2015,Forero_Romero_2014}. 
In particular, \citet{Codis2012} find that the orientation of a halo's spin 
relative to the filament in which it resides has a mass dependence with a turnover 
from pointing parallel to the filament to perpendicular 
at a critical mass of $\sim$ 5($\pm$1) $\times$ $10^{12}\msun$. 
This mass is greater than that of our Milky way-mass halos and considerably 
lower than that of our cluster-mass halos. This suggests that Milky Way-mass 
halos will be oriented differently relative to the present-day
large-scale structure than cluster-mass halos will. This is interesting to 
note and may be important for a variety of forthcoming applications, 
including gravitational lensing studies, but we will not pursue this 
further in this work. 

\subsection{The Distribution of Subhalo Mass as a Function of Viewing Angle}

In order to interpret strong gravitational lensing measurements, it is most 
practical to focus on the mass in the central region of the halo and  
specifically, the mass within approximately the Einstein radius in projection.
For lenses at z=0.5 composed of a galaxy within a dark matter halo within the mass range of our Milky Way-mass halos, with a source at z=2, we estimate an Einstein radius of approximately 2 arcsecs or 12.2 kpc and for lenses composed of a galaxy in a cluster-mass dark matter halo they are approximately 10 arcsecs or 61 kpc.
Because of the ellipsoidal shape of halos, the projected matter density near 
the center will vary depending on the orientation from which the halo is viewed 
and a greater projected matter density will be observed near the center when the halo is viewed 
along its long axis compared to its shortest.
We predict that a greater projected density of subhalos at 
the halo center will also be observed when the halo is viewed along 
this axis.

\begin{figure*}
    \centering
    \begin{subfigure}[b]{0.47\textwidth}
        \includegraphics[width=\linewidth]{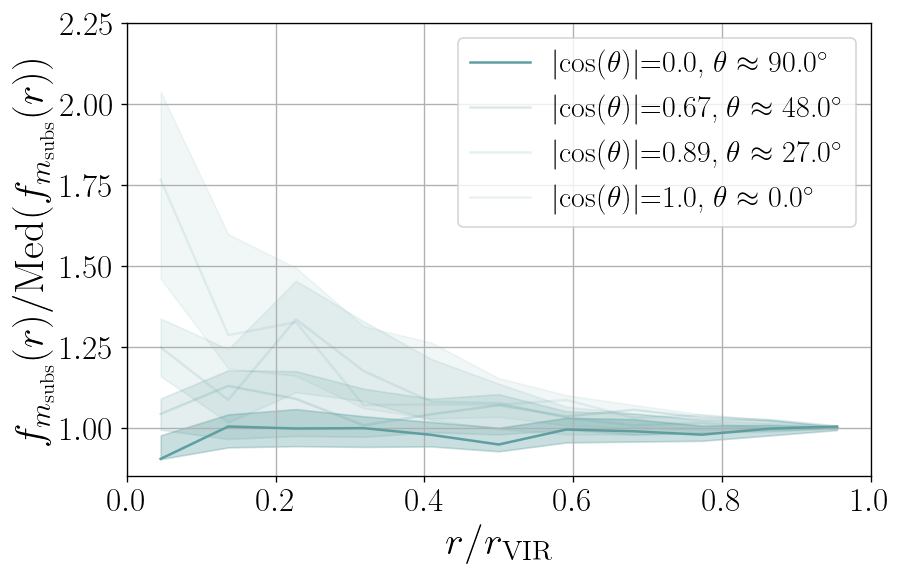}
        \caption{}
    \end{subfigure}
    \begin{subfigure}[b]{0.47\textwidth}
        \includegraphics[width=\linewidth]{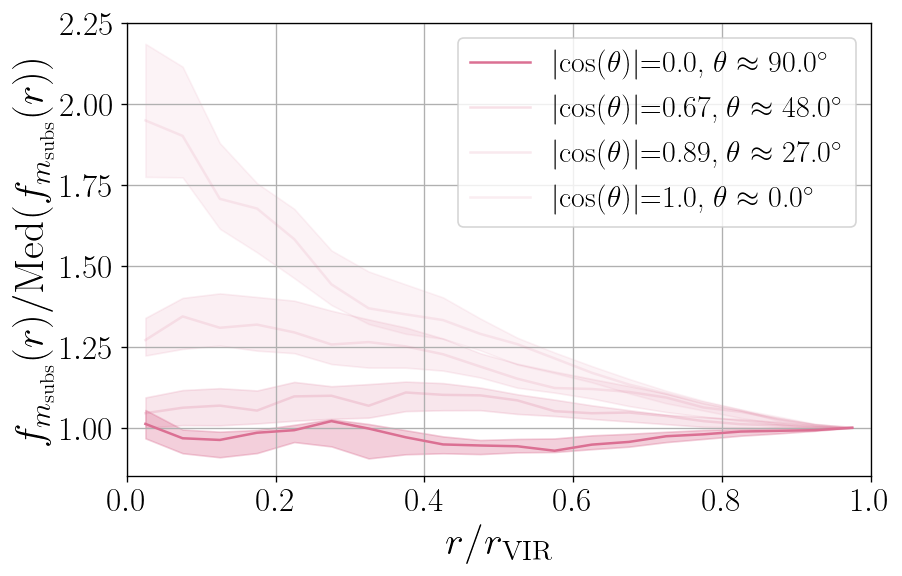}
        \caption{}
    \end{subfigure}
    \begin{subfigure}[b]{0.47\textwidth}
        \includegraphics[width=\linewidth]{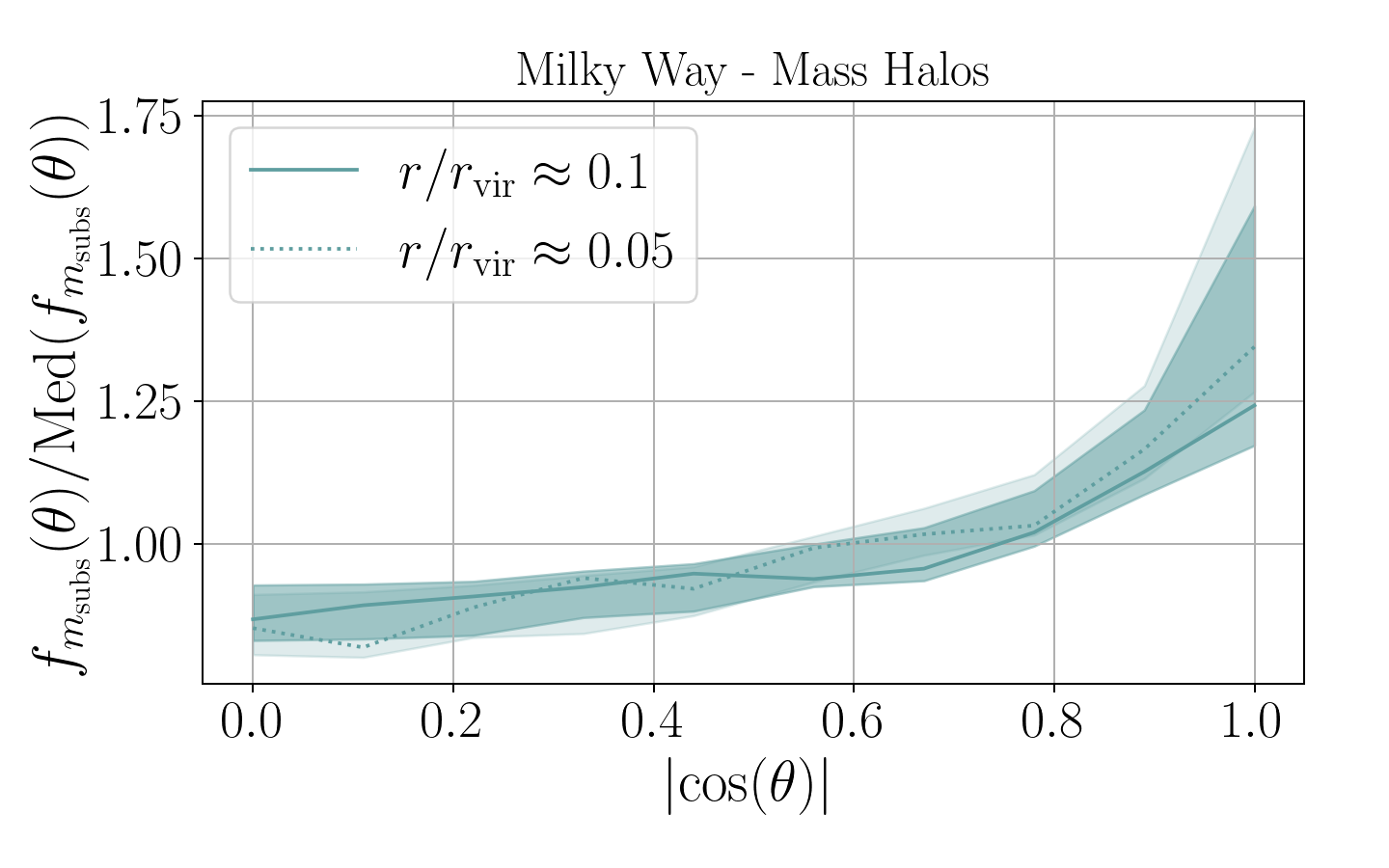}
        \caption{}
    \end{subfigure}
    \begin{subfigure}[b]{0.47\textwidth}
        \includegraphics[width=\linewidth]{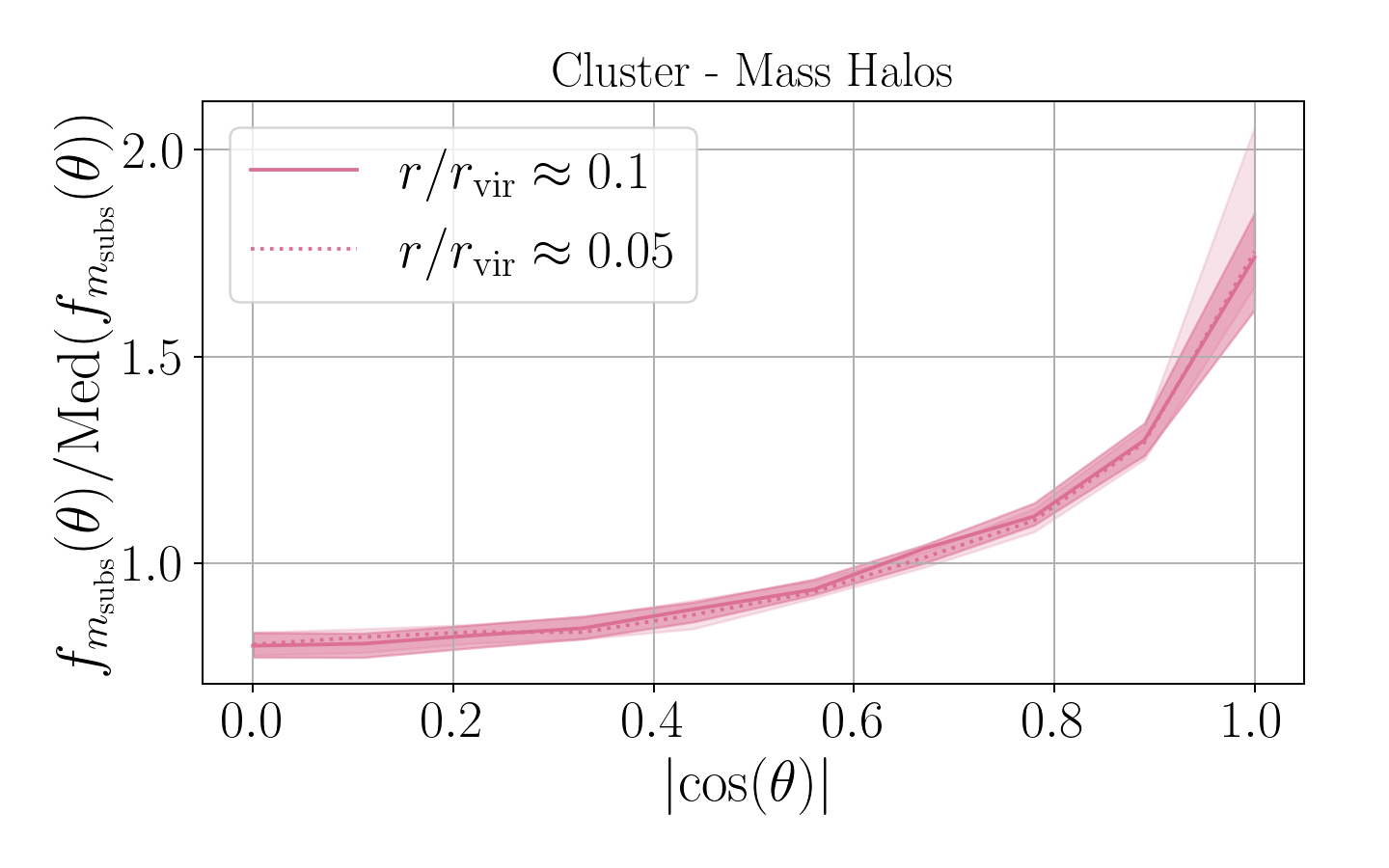}
        \caption{}
    \end{subfigure}
    \caption{The top two panels show subhalo mass fraction captured within cylinders projected along predefined axes on a uniformly spaced grid on the surface of a sphere normalized by the median mass fraction across all halos and angles. The left panel in blue corresponds to Milky Way-mass halos and the right panel in pink to cluster-mass halos. The x-axis is the ratio of the radius of the cylinder to virial radius of the halo, and, the y-axis is the normalized subhalo mass fraction. The solid lines are the median normalized mass fraction within bins of $r/r_{\rm{VIR}}$ at four different angular separations ranging from aligned with the major axis (lightest shaded line) to perpendicular to the major axis (darkest shaded line). The bottom two panels contain the same information presented in a different way. For these figures, the x-axis is the angular separation. The solid and dotted lines correspond to the mass fraction within cylinders of radii equal to approximately 10\% and 5\% of the virial radius. The shaded regions are the error margins corresponding to 14th and 86th percentiles.}
    \label{figure:angle_v_mass_frac}
\end{figure*}

Determining the influence of subhalos on strong lensing systems is a topic of great contemporary interest.
We have already demonstrated that subhalos are distributed anisotropically 
around their hosts and that subhalos are found 
preferentially aligned the long axes of their hosts.
We quantify this effect to inform its importance in previous and forthcoming lensing measurements.
We show the results from this analysis in \autoref{figure:angle_v_mass_frac} 
where the left panels correspond to Milky Way-mass halos and the right panels to cluster-mass halos.
The panels of \autoref{figure:angle_v_mass_frac} depict the mass fraction 
in subhalos in a cylindrical projection as a function of both radius of the 
cylindrical projection (top panels) and viewing angle 
relative to the major axis of the halo (bottom panels). 
The cylinder scheme was discussed in \autoref{Section:axis_calcs} 
and it may be useful to review \autoref{fig:cylinders} to visualize the geometry 
under consideration. In all panels, the vertical axis is the ratio of the 
projected subhalo mass fraction enclosed within a particular 
cylinder, oriented at an angle $\theta$ away from the halo major axis, 
divided by the median mass fraction for all 
cylindrical projections across all halos and angles.

In the top two panels of \autoref{figure:angle_v_mass_frac}, 
the solid line is the median normalized subhalo mass fraction within bins of $r/r_{\rm{VIR}}$ at four different angular separations ranging from aligned with the major axis (lightest shaded line) to perpendicular to the major axis (darkest shaded line).  The shaded regions are the error margins corresponding to 14th and 86th percentiles.
Mass fractions are normalized by the median subhalo mass fraction across all halos and angles. 
Consequently, if the subhalos were distributed isotropically, the solid lines in each panel 
would be horizontal lines with a value of 1.0. 
In both Milky Way-mass and cluster-mass halos, 
the subhalo mass fraction increases as cylinder orientation approaches the major axis. 
Smaller cylinder radii lead to dramatically larger differences, which makes sense because 
smaller cylinder radii correspond to averaging over a smaller physical scale so that correlations 
are not washed out by subhalos that are poorly aligned with the major axis, but fall within the 
cylinder radius. This effect is more dramatic for cluster-mass halos where the mass fraction is approximately 1.95 times greater at small cylinder radii along the major axis compared to perpendicular, whereas the trend for Milky Way-mass halos is both smaller (1.75 times greater at small radii along the major axis) and noisier (due to the smaller sample size).
The bottom two panels show the median normalized mass fraction within bins of separation angle at two cylinder radii equal to approximately 5\% and 10\% of the virial radius.
Once again, the shaded regions are the bootstrapped error margins corresponding to 14th and 86th percentiles.
In these figures we again note an increase in mass fraction as cylinder orientation approaches the major axis for the same reason as discussed above. We note that the difference in maximum mass fraction between the top and bottom figures is due to the fact that in the bottom figures we are binning data in cos($\theta$).

This result naturally emerges from the anisotropies already 
discussed in \autoref{sub_axis_alignment}. 
We saw overall stronger alignments with particular axes 
for cluster-mass halos in \autoref{sub_axis_alignment} as well, 
so the difference in the strength of this effect between 
Milky Way-mass halos and cluster-mass halos is not surprising. 
We attribute the noisiness in the Milky Way result to the fact 
that there are fewer than half as many Milky Way-mass host halos 
compared to cluster-mass host halos. Moreover, smaller host halos 
tend to have a smaller total fraction of their mass in subhalos, 
so the Milky Way-mass hosts contain fewer subhalos and thus there is 
a larger shot noise associated with subhalo position.

For lensing, a massive subhalo of some mass, $M$, 
in the line of sight will leave a larger lensing 
signal than a collection of smaller subhalos with collective mass, $M$. 
To this end, we ran a smaller scale version of the analysis discussed above with a variety 
of subsamples computed using threshold samples of subhalos above a minimum 
mass. We calculate projected subhalo mass for subhalos of mass, 
$M$, greater than 0.025, 0.05, 0.075, and 0.1 times the total halo virial mass. 
These numbers are selected such that they fall within the range of available 
subhalo masses -- Milky Way-mass halos in our sample do not have 
subhalos more massive than 0.2 times their total halo virial mass and 
only a handful of cluster-mass halos have subhalos that exceed this threshold.
For this iteration of the analysis,
we calculate the mass fraction at only two radii on the scale of the expected Einstein radius,
0.05 and 0.1 times the halo virial radius, along 500 randomly selected axes through the halo.
We find that as this lower mass limit is increased, 
there is a small increase in the overabundance of projected subhalo 
mass near the major axis, although, the change is not statistically 
significant. When considered together with our other results indicating 
subhalo anisotropies, we nevertheless encourage the reader 
to be mindful of this potential effect.

\section{Conclusion and Discussion}
\label{sec:discussion_conclusions}

\subsection{Conclusion}
\label{sec:conclusions}
In this paper we have shown that subhalos are distributed anisotropically with respect to their host halos.
In particular, 
subhalos tend to reside preferentially near the major axis of their host halo.
We examined specifically the following relationships between the 
host halo and its subhalo population: 
(1) the degree to which the angular 
momentum of the host halo is correlated with the 
net angular momentum of its associated subhalo population; 
(2) the alignment of the principle axes of the host halo with 
the principle axes defined by its subhalo population; 
and (3) we measured the projected mass in subhalos 
as a function of projection angle relative to the major 
axis of the host. This study anticipates applying these results 
to the interpretation of strong lensing observations, but we 
have phrased the alignment in a variety of ways so that 
these alignment results may be applied more broadly in a 
variety of other applications. To our knowledge, 
this is the only analysis of this type performed using 
zoom-in simulations of comparably high resolution.
This has enabled us to study subhalo-host relationships in greater detail 
than previous studies. Our main results are as follows. 

\begin{itemize}
    \item Both cluster-mass and Milky Way-mass host halo's major axes are generally well aligned with those of their subhalos. In this broad sense, subhalos are preferentially oriented in the same way as the overall mass distribution of their host halos. For Milky Way-mass halos, the median angle between the host halo major axis the subhalo component major axis is $\theta$ = $30\pm 4^{\circ}$ or $\vert \cos(\theta)\vert$ = 0.86 $\pm$ 0.04. For Cluster-mass halos, this median angle is $\theta$ = $25\pm 3.5^{\circ}$ or $\vert \cos(\theta)\vert$ = 0.91 $\pm$ 0.03.\\
    \item Both cluster-mass and Milky Way-mass host halo's angular momenta are fairly well aligned with the net angular momenta of their subhalos. For Milky Way-mass halos, the median angle between the host halo angular momentum the net subhalo angular momentum is $\theta$ = $50\pm 9^{\circ}$ or cos($\theta$) = 0.63 $\pm$ 0.12. For Cluster-mass halos, this median angle is $\theta$ = $56\pm 5.5^{\circ}$ or cos($\theta$) = 0.56 $\pm$ 0.08.\\
    \item Halos that are viewed along their major axis have a greater projected subhalo mass near their centers compared to those viewed along a random axis direction. As seen in the top panels of \autoref{figure:angle_v_mass_frac}, within projected radii 
    of a few percent of the virial radius of the host halo, the fraction of mass in subhalos is $\sim 175\%$ larger for Milky Way mass halos and as much as $\sim 195\%$ larger for cluster halos when projected along the major axis as compared to the average from a random projection. To within 
    our ability to measure, this preference does not have a strong dependence upon the 
    mass of the subhalo.\\
\end{itemize}

\subsection{Discussion}

\label{sec:Discussion}

The intrinsic alignment between halo major axes provides insight into structure growth and halo formation.
Within hierarchical formation models, overdense regions of the density field collapse into sheets and filaments and underdense regions into voids.
Dark matter halos form at peaks of the density field which collapse and grow over time through mergers and smooth accretion.

Broadly speaking, the alignments that we have explored can be understood 
within the standard paradigm of hierarchical structure formation. 
Early-forming halos collapse from peaks in the initial density field and form 
triaxial structures \citep{zeldovich1970}. 
Tidal torquing then produces net halo angular momenta that are roughly 
orthogonal to halo major axes \citep{peebles1969,white1984}. 
Those halos which reside within larger-scale overdensities then merge to form larger halos 
and so the population of massive halos is built from mergers with smaller halos. 
In the net, these mergers take place along strongly correlated directions \citep{zentner2005, Libeskind_2007,Libeskind2010}.
In particular, halos preferentially merge along the filamentary structure.
Sequences of mergers from preferential directions cause halos to become 
elongated in the direction of the merger. As a result, host halos tend to be prolate and their major 
axes tend to point along the filaments \citep{Jing2002,Bailin_2005,Faltenbacher2005,zentner2005}. 
The very same mergers supply host halos 
with their subhalos and the subhalos tend to continue on elongated orbits that reflect 
the direction from which they merged \citep{zentner2005}. Of course, 
the details of these correlations must be computed from numerical simulations 
and many of the papers cited in this paragraph have done just that.

This overall picture is supported by observational studies. Halo ellipticity has 
now been detected via weak gravitational lensing and those lensing measurements 
also confirm that the major axes of galaxies must be correlated with the 
major axes of their host halos. The non-zero ellipticities of halos have been 
detected in both weak lensing 
\citep[e.g.,][]{evans_bridle2009,clampitt_jain2016,gonzalez+2021,robison+2023} 
and strong lensing \citep[e.g.,][]{okabe+2020}. Moreover, these 
measurements are broadly consistent with theoretical expectations 
\citep[e.g.,][]{bett2012,velliscig+2015,chisari+2017}. Of particular 
interest in the current context are the results of \citet{evans_bridle2009} 
and \citet{gonzalez+2021} who found that the orientation of the dark matter 
ellipse was correlated with the distribution of member galaxies in clusters. 
This suggests a detection of the bias for finding subhalos to be aligned with 
the major axes of their host halos. There are also a number of 
studies of SDSS data that show that satellite galaxies tend to reside along 
the major axes of their host galaxies \citep{Brainerd_2005,Yang_2006,Wang_2018}.

The alignments between host dark matter halos and their subhalos has 
implications for a variety of observations and particularly for the analysis 
of strong gravitational lensing. Strong gravitational lensing has become an 
indispensable probe of galaxy and halo structure. Among the most 
ambitious achievements and aims of strong lensing is to identify 
lensing features caused by subhalos within a primary lens system 
and thereby constrain the properties of the dark matter. This can 
be done using flux-ratio anomalies 
\citep{Mao_1998,Metcalf_2001,Dalal_2002,Brada_2002,Keeton_2003,Metcalf_2002,Metcalf_2004,Treu_2016,Gilman_2018,Harvey_2019,Treu_2022,zelko_2023,keeley+2024}
and perturbations in lensing arcs \citep{Meneghetti2006,Vegetti2010,despali2018,Minor2021}. 
In each case, it is necessary to know the amount of substructure one should 
expect near the Einstein radius of the lens systems as viewed in projection 
along the line of sight.

We found that the number of subhalos as well as the projected mass 
fraction in subhalos is greater when viewed along a two-dimensional 
projection aligned with the major axis of the host halo. The mass 
fraction in subhalos declines as the viewing angle with respect to 
the principle axis increases. The halos themselves are typically 
nearly prolate ellipsoids, so the total mass density in projection 
is higher when viewed along a direction closely aligned with 
the major axis. This suggests that halos may be more likely 
to serve as strong lenses when viewed at an angle nearly aligned with 
the major axis, though the strength of this preference depends 
upon the poorly constrained degree of alignment between the 
principle axes of a lens halo and the principle axes of the galaxy 
it hosts. Indeed, \citet{Hennawi2007} estimate that the 
mean value of the cosine of the viewing angle with respect to the 
major axis, $\vert \cos(\theta)\vert$, should be $\sim 0.67$ for 
multiply-imaged systems, rather than $0.5$, as would be expected if 
lensing probability were independent of viewing angle. There is a 
clear implication of these results. If halos that serve as 
strong lenses are observed in projections that are nearly 
projections along the major axis, at least preferentially, then 
the expected mass fraction in subhalos should be substantially 
larger than predictions for the global mass fraction in subhalos. 
In follow-up work, we will explore the degree to which this 
potential bias can affect the interpretation of strong 
lensing observations and we advise those analyzing strong 
lensing data to be aware of this potential bias.

There are possibly other important consequences of the alignments highlighted in this work. Our finding that subhalo and host halo angular momenta are aligned 
has consequences for galaxy formation studies. 
For example, it is common practice for semi-analytic galaxy formation models 
to adopt host halo angular momentum as a proxy for galaxy angular momentum either probabilistically \citep{Somerville2008,Guo2011} or deterministically \citep{Benson2012}. Interestingly, our results suggest that a large fraction of the net angular momentum of the host halo is carried by the subhalos reside on orbits that are close to the plane of any galactic disk. Sufficiently large subhalos will likely host their own galaxies.
This provides an interesting contrast to a now well-known observation in the Local Group.  
The dwarf satellite galaxies of the Milky Way seem to orbit the Milky Way in a plane that is 
nearly \textit{perpendicular} to the galaxy disk so that the angular momentum carried by 
the dwarf satellites is nearly perpendicular to that angular momentum of the Milky Way galaxy \citep{kroupa_2005,Pawlowski2018}. Satellites of Andromeda are found to be distributed anisotropically \citep{DolivaDolinsky2023} and orbit the disk of Andromeda in a separate plane at a tilt of 
$\sim$50$^\circ$ \citep{Pawlowski2018} with respect to the angular momentum of Andromeda.
Such planar orbits of satellites have also been found in simulations using Milky Way and 
Andromeda galaxy analogs in \citet{SantosSantos2020} and \citet{Samuel2021} with the latter showing that they can exist in a plane perpendicular to the galactic disk.
However, recent work indicates that the perpendicular alignment seen in the Milky Way 
could be temporary and due to a chance alignment with its two most distant satellites 
(Leo I and II) \citep{Sales2023}. Additionally, 
prior simulation work by \citet{fielder2018} found host halo angular momentum 
to be the weakest indicator of subhalo abundance, further reinforcing 
the idea of alignment between host halo and subhalo angular momenta.
All things considered, we believe that caution must be taken 
by galaxy formation studies when defining galaxy angular momentum.

In this paper, we provide a comprehensive look at the biased distribution of 
subhalos about their host halos. Using zoom-in simulations, we have shown that subhalos 
are distributed anisotropically with respect to their host halos. We find that 
(i) they are preferentially located along the major axis of their host halo and 
(ii) they orbit in a plane perpendicular to the spin axis of their host halo. 
These are effects that may be potentially important in galaxy formation models 
and, in particular, in the interpretation of strong lens systems.

\section*{Acknowledgements}

We thank Atınç Çağan Şengül for insightful comments and helpful discussion that improved this manuscript.

This research made use of Python, along with many community-developed or maintained software packages, including
IPython \citep{ipython},
Jupyter (\http{jupyter.org}),
Matplotlib \citep{matplotlib},
NumPy \citep{numpy},
Pandas \citep{pandas},
and SciPy \citep{scipy}.
This research made use of NASA's Astrophysics Data System for bibliographic information.

This work used data from the Symphony suite of simulations (http://web.stanford.edu/group/gfc/symphony/), which was supported by the Kavli Institute for Particle Astrophysics and Cosmology at Stanford University and SLAC National Accelerator Laboratory, and by the U.S. Department of Energy under contract number DE-AC02-76SF00515 to SLAC National Accelerator Laboratory.

\section*{Data Availability}
The Symphony data products used in this work are publicly available at \https{phil-mansfield.github.io/symphony}.

\bibliographystyle{mnras}
\bibliography{refs,software}

\begin{thebibliography}{}
\makeatletter
\relax
\def\mn@urlcharsother{\let\do\@makeother \do\$\do\&\do\#\do\^\do\_\do\%\do\~}
\def\mn@doi{\begingroup\mn@urlcharsother \@ifnextchar [ {\mn@doi@} {\mn@doi@[]}}
\def\mn@doi@[#1]#2{\def\@tempa{#1}\ifx\@tempa\@empty \href {http://dx.doi.org/#2} {doi:#2}\else \href {http://dx.doi.org/#2} {#1}\fi \endgroup}
\def\mn@eprint#1#2{\mn@eprint@#1:#2::\@nil}
\def\mn@eprint@arXiv#1{\href {http://arxiv.org/abs/#1} {{\tt arXiv:#1}}}
\def\mn@eprint@dblp#1{\href {http://dblp.uni-trier.de/rec/bibtex/#1.xml} {dblp:#1}}
\def\mn@eprint@#1:#2:#3:#4\@nil{\def\@tempa {#1}\def\@tempb {#2}\def\@tempc {#3}\ifx \@tempc \@empty \let \@tempc \@tempb \let \@tempb \@tempa \fi \ifx \@tempb \@empty \def\@tempb {arXiv}\fi \@ifundefined {mn@eprint@\@tempb}{\@tempb:\@tempc}{\expandafter \expandafter \csname mn@eprint@\@tempb\endcsname \expandafter{\@tempc}}}

\bibitem[\protect\citeauthoryear{Alard}{Alard}{2008}]{Alard2008}
Alard C.,  2008, \mn@doi [Monthly Notices of the Royal Astronomical Society] {10.1111/j.1365-2966.2008.13397.x}, 388, 375

\bibitem[\protect\citeauthoryear{Arag{\'o}n-Calvo, van~de Weygaert, Jones  \& van~der Hulst}{Arag{\'o}n-Calvo et~al.}{2007}]{Arag_n_Calvo_2007}
Arag{\'o}n-Calvo M.~A.,  van~de Weygaert R.,  Jones B. J.~T.,   van~der Hulst J.~M.,  2007, \mn@doi [The Astrophysical Journal] {10.1086/511633}, 655, L5

\bibitem[\protect\citeauthoryear{Aubert, Pichon  \& Colombi}{Aubert et~al.}{2004}]{Aubert_2004}
Aubert D.,  Pichon C.,   Colombi S.,  2004, \mn@doi [Monthly Notices of the Royal Astronomical Society] {10.1111/j.1365-2966.2004.07883.x}, 352, 376–398

\bibitem[\protect\citeauthoryear{{Azzaro}, {Zentner}, {Prada}  \& {Klypin}}{{Azzaro} et~al.}{2006}]{Azzaro+06}
{Azzaro} M.,  {Zentner} A.~R.,  {Prada} F.,   {Klypin} A.~A.,  2006, \mn@doi [\apj] {10.1086/499262}, \href {https://ui.adsabs.harvard.edu/abs/2006ApJ...645..228A} {645, 228}

\bibitem[\protect\citeauthoryear{{Azzaro}, {Patiri}, {Prada}  \& {Zentner}}{{Azzaro} et~al.}{2007}]{Azzaro+07}
{Azzaro} M.,  {Patiri} S.~G.,  {Prada} F.,   {Zentner} A.~R.,  2007, \mn@doi [\mnras] {10.1111/j.1745-3933.2007.00282.x}, \href {https://ui.adsabs.harvard.edu/abs/2007MNRAS.376L..43A} {376, L43}

\bibitem[\protect\citeauthoryear{Bailin \& Steinmetz}{Bailin \& Steinmetz}{2005}]{Bailin_2005}
Bailin J.,  Steinmetz M.,  2005, \mn@doi [The Astrophysical Journal] {10.1086/430397}, 627, 647

\bibitem[\protect\citeauthoryear{{Barnes} \& {Efstathiou}}{{Barnes} \& {Efstathiou}}{1987}]{barnes1987}
{Barnes} J.,  {Efstathiou} G.,  1987, \mn@doi [\apj] {10.1086/165480}, \href {https://ui.adsabs.harvard.edu/#abs/1987ApJ...319..575B} {319, 575}

\bibitem[\protect\citeauthoryear{{Becker}}{{Becker}}{2015}]{becker2015}
{Becker} M.~R.,  2015, preprint, \href {https://ui.adsabs.harvard.edu/#abs/2015arXiv150703605B} {p. arXiv:1507.03605} (\mn@eprint {arXiv} {1507.03605})

\bibitem[\protect\citeauthoryear{{Behroozi}, {Wechsler}  \& {Wu}}{{Behroozi} et~al.}{2013}]{behroozi2013}
{Behroozi} P.~S.,  {Wechsler} R.~H.,   {Wu} H.-Y.,  2013, \mn@doi [\apj] {10.1088/0004-637X/762/2/109}, \href {https://ui.adsabs.harvard.edu/#abs/2013ApJ...762..109B} {762, 109}

\bibitem[\protect\citeauthoryear{Benson}{Benson}{2012}]{Benson2012}
Benson A.~J.,  2012, \mn@doi [New Astronomy] {10.1016/j.newast.2011.07.004}, 17, 175

\bibitem[\protect\citeauthoryear{{Bett}}{{Bett}}{2012}]{bett2012}
{Bett} P.,  2012, \mn@doi [\mnras] {10.1111/j.1365-2966.2011.20258.x}, \href {https://ui.adsabs.harvard.edu/abs/2012MNRAS.420.3303B} {420, 3303}

\bibitem[\protect\citeauthoryear{{Blumenthal}, {Faber}, {Primack}  \& {Rees}}{{Blumenthal} et~al.}{1984}]{blumenthal1984}
{Blumenthal} G.~R.,  {Faber} S.~M.,  {Primack} J.~R.,   {Rees} M.~J.,  1984, \mn@doi [\nat] {10.1038/311517a0}, \href {https://ui.adsabs.harvard.edu/abs/1984Natur.311..517B} {311, 517}

\bibitem[\protect\citeauthoryear{{Bode}, {Ostriker}  \& {Turok}}{{Bode} et~al.}{2001}]{bode+2001}
{Bode} P.,  {Ostriker} J.~P.,   {Turok} N.,  2001, \mn@doi [\apj] {10.1086/321541}, \href {https://ui.adsabs.harvard.edu/abs/2001ApJ...556...93B} {556, 93}

\bibitem[\protect\citeauthoryear{Bradač, Schneider, Steinmetz, Lombardi, King  \& Porcas}{Bradač et~al.}{2002}]{Brada_2002}
Bradač M.,  Schneider P.,  Steinmetz M.,  Lombardi M.,  King L.~J.,   Porcas R.,  2002, \mn@doi [Astronomy &amp; Astrophysics] {10.1051/0004-6361:20020559}, 388, 373–382

\bibitem[\protect\citeauthoryear{Brainerd}{Brainerd}{2005}]{Brainerd_2005}
Brainerd T.~G.,  2005, \mn@doi [The Astrophysical Journal] {10.1086/432713}, 628, L101–L104

\bibitem[\protect\citeauthoryear{{Brainerd} \& {Samuels}}{{Brainerd} \& {Samuels}}{2021}]{brainerd2021}
{Brainerd} T.,  {Samuels} A.,  2021, in American Astronomical Society Meeting Abstracts. p. 156.02

\bibitem[\protect\citeauthoryear{{Bullock}}{{Bullock}}{2010}]{bullock2010}
{Bullock} J.~S.,  2010, preprint, \href {https://ui.adsabs.harvard.edu/#abs/2010arXiv1009.4505B} {p. arXiv:1009.4505} (\mn@eprint {arXiv} {1009.4505})

\bibitem[\protect\citeauthoryear{Chiba \& Takahashi}{Chiba \& Takahashi}{2002}]{Chiba_2002}
Chiba T.,  Takahashi R.,  2002, \mn@doi [Progress of Theoretical Physics] {10.1143/ptp.107.625}, 107, 625–630

\bibitem[\protect\citeauthoryear{{Chisari} et~al.,}{{Chisari} et~al.}{2017}]{chisari+2017}
{Chisari} N.~E.,  et~al., 2017, \mn@doi [\mnras] {10.1093/mnras/stx1998}, \href {https://ui.adsabs.harvard.edu/abs/2017MNRAS.472.1163C} {472, 1163}

\bibitem[\protect\citeauthoryear{{Clampitt} \& {Jain}}{{Clampitt} \& {Jain}}{2016}]{clampitt_jain2016}
{Clampitt} J.,  {Jain} B.,  2016, \mn@doi [\mnras] {10.1093/mnras/stw254}, \href {https://ui.adsabs.harvard.edu/abs/2016MNRAS.457.4135C} {457, 4135}

\bibitem[\protect\citeauthoryear{Codis, Pichon, Devriendt, Slyz, Pogosyan, Dubois  \& Sousbie}{Codis et~al.}{2012}]{Codis2012}
Codis S.,  Pichon C.,  Devriendt J.,  Slyz A.,  Pogosyan D.,  Dubois Y.,   Sousbie T.,  2012, \mn@doi [Monthly Notices of the Royal Astronomical Society] {10.1111/j.1365-2966.2012.21636.x}, 427, 3320

\bibitem[\protect\citeauthoryear{{Col{\'\i}n}, {Avila-Reese}  \& {Valenzuela}}{{Col{\'\i}n} et~al.}{2000}]{colin+2000}
{Col{\'\i}n} P.,  {Avila-Reese} V.,   {Valenzuela} O.,  2000, \mn@doi [\apj] {10.1086/317057}, \href {https://ui.adsabs.harvard.edu/abs/2000ApJ...542..622C} {542, 622}

\bibitem[\protect\citeauthoryear{{Conn} et~al.,}{{Conn} et~al.}{2013}]{conn_2013}
{Conn} A.~R.,  et~al., 2013, \mn@doi [\apj] {10.1088/0004-637X/766/2/120}, \href {https://ui.adsabs.harvard.edu/abs/2013ApJ...766..120C} {766, 120}

\bibitem[\protect\citeauthoryear{{D'Onghia}, {Burkert}, {Murante}  \& {Khochfar}}{{D'Onghia} et~al.}{2006}]{D_Onghia2006}
{D'Onghia} E.,  {Burkert} A.,  {Murante} G.,   {Khochfar} S.,  2006, \mn@doi [\mnras] {10.1111/j.1365-2966.2006.10996.x}, \href {https://ui.adsabs.harvard.edu/abs/2006MNRAS.372.1525D} {372, 1525}

\bibitem[\protect\citeauthoryear{Dalal \& Kochanek}{Dalal \& Kochanek}{2002}]{Dalal_2002}
Dalal N.,  Kochanek C.~S.,  2002, \mn@doi [The Astrophysical Journal] {10.1086/340303}, 572, 25–33

\bibitem[\protect\citeauthoryear{Deason et~al.,}{Deason et~al.}{2011}]{Deason_2011}
Deason A.~J.,  et~al., 2011, \mn@doi [Monthly Notices of the Royal Astronomical Society] {10.1111/j.1365-2966.2011.18884.x}, 415, 2607–2625

\bibitem[\protect\citeauthoryear{Despali, Vegetti, White, Giocoli  \& van~den Bosch}{Despali et~al.}{2018}]{despali2018}
Despali G.,  Vegetti S.,  White S. D.~M.,  Giocoli C.,   van~den Bosch F.~C.,  2018, \mn@doi [Monthly Notices of the Royal Astronomical Society] {10.1093/mnras/sty159}, 475, 5424

\bibitem[\protect\citeauthoryear{Doliva-Dolinsky et~al.,}{Doliva-Dolinsky et~al.}{2023}]{DolivaDolinsky2023}
Doliva-Dolinsky A.,  et~al., 2023, \mn@doi [The Astrophysical Journal] {10.3847/1538-4357/acdcf6}, 952, 72

\bibitem[\protect\citeauthoryear{{Dubinski}}{{Dubinski}}{1994}]{Dubinski_1994}
{Dubinski} J.,  1994, \mn@doi [\apj] {10.1086/174512}, \href {https://ui.adsabs.harvard.edu/abs/1994ApJ...431..617D} {431, 617}

\bibitem[\protect\citeauthoryear{{Dubinski} \& {Carlberg}}{{Dubinski} \& {Carlberg}}{1991}]{dubinkski_1991}
{Dubinski} J.,  {Carlberg} R.~G.,  1991, \mn@doi [\apj] {10.1086/170451}, \href {https://ui.adsabs.harvard.edu/abs/1991ApJ...378..496D} {378, 496}

\bibitem[\protect\citeauthoryear{{Evans} \& {Bridle}}{{Evans} \& {Bridle}}{2009}]{evans_bridle2009}
{Evans} A. K.~D.,  {Bridle} S.,  2009, \mn@doi [\apj] {10.1088/0004-637X/695/2/1446}, \href {https://ui.adsabs.harvard.edu/abs/2009ApJ...695.1446E} {695, 1446}

\bibitem[\protect\citeauthoryear{Faltenbacher, Gottlöber, Kerscher  \& Müller}{Faltenbacher et~al.}{2002}]{Faltenbacher_2002}
Faltenbacher A.,  Gottlöber S.,  Kerscher M.,   Müller V.,  2002, \mn@doi [Astronomy {\&} Astrophysics] {10.1051/0004-6361:20021263}, 395, 1

\bibitem[\protect\citeauthoryear{Faltenbacher, Allgood, Gottlöber, Yepes  \& Hoffman}{Faltenbacher et~al.}{2005}]{Faltenbacher2005}
Faltenbacher A.,  Allgood B.,  Gottlöber S.,  Yepes G.,   Hoffman Y.,  2005, \mn@doi [Monthly Notices of the Royal Astronomical Society] {10.1111/j.1365-2966.2005.09386.x}, 362, 1099

\bibitem[\protect\citeauthoryear{{Fielder}, {Mao}, {Newman}, {Zentner}  \& {Licquia}}{{Fielder} et~al.}{2018}]{fielder2018}
{Fielder} C.~E.,  {Mao} Y.-Y.,  {Newman} J.~A.,  {Zentner} A.~R.,   {Licquia} T.~C.,  2018, preprint, \href {http://adsabs.harvard.edu/abs/2018arXiv180705180F} {} (\mn@eprint {arXiv} {1807.05180})

\bibitem[\protect\citeauthoryear{Fielder, Mao, Zentner, Newman, Wu  \& Wechsler}{Fielder et~al.}{2020}]{fielder2020}
Fielder C.~E.,  Mao Y.-Y.,  Zentner A.~R.,  Newman J.~A.,  Wu H.-Y.,   Wechsler R.~H.,  2020, \mn@doi [Monthly Notices of the Royal Astronomical Society] {10.1093/mnras/staa2851}, 499

\bibitem[\protect\citeauthoryear{Forero-Romero, Contreras  \& Padilla}{Forero-Romero et~al.}{2014}]{Forero_Romero_2014}
Forero-Romero J.~E.,  Contreras S.,   Padilla N.,  2014, \mn@doi [Monthly Notices of the Royal Astronomical Society] {10.1093/mnras/stu1150}, 443, 1090–1102

\bibitem[\protect\citeauthoryear{Gilman, Birrer, Treu, Keeton  \& Nierenberg}{Gilman et~al.}{2018}]{Gilman_2018}
Gilman D.,  Birrer S.,  Treu T.,  Keeton C.~R.,   Nierenberg A.,  2018, \mn@doi [Monthly Notices of the Royal Astronomical Society] {10.1093/mnras/sty2261}, 481, 819–834

\bibitem[\protect\citeauthoryear{{Gonzalez}, {Makler}, {Garc{\'\i}a Lambas}, {Chalela}, {Pereira}, {Van Waerbeke}, {Shan}  \& {Erben}}{{Gonzalez} et~al.}{2021}]{gonzalez+2021}
{Gonzalez} E.~J.,  {Makler} M.,  {Garc{\'\i}a Lambas} D.,  {Chalela} M.,  {Pereira} M. E.~S.,  {Van Waerbeke} L.,  {Shan} H.,   {Erben} T.,  2021, \mn@doi [\mnras] {10.1093/mnras/staa3570}, \href {https://ui.adsabs.harvard.edu/abs/2021MNRAS.501.5239G} {501, 5239}

\bibitem[\protect\citeauthoryear{Grebel}{Grebel}{1998}]{grebel_1998}
Grebel E.~K.,  1998, Evolutionary Histories of Dwarf Galaxies in the Local Group (\mn@eprint {arXiv} {astro-ph/9812443})

\bibitem[\protect\citeauthoryear{Guo et~al.,}{Guo et~al.}{2011}]{Guo2011}
Guo Q.,  et~al., 2011, \mn@doi [Monthly Notices of the Royal Astronomical Society] {10.1111/j.1365-2966.2010.18114.x}, 413, 101

\bibitem[\protect\citeauthoryear{Han, Wang, Avestruz  \& Anbajagane}{Han et~al.}{2023}]{han2023}
Han C.,  Wang K.,  Avestruz C.,   Anbajagane D.,  2023, Subhalos in Galaxy Clusters: Coherent Accretion and Internal Orbits (\mn@eprint {arXiv} {2312.08337})

\bibitem[\protect\citeauthoryear{Hartwick}{Hartwick}{2000}]{Hartwick_2000}
Hartwick F. D.~A.,  2000, \mn@doi [The Astronomical Journal] {10.1086/301332}, 119, 2248–2253

\bibitem[\protect\citeauthoryear{Harvey, Valkenburg, Tamone, Boyarsky, Courbin  \& Lovell}{Harvey et~al.}{2019}]{Harvey_2019}
Harvey D.,  Valkenburg W.,  Tamone A.,  Boyarsky A.,  Courbin F.,   Lovell M.,  2019, \mn@doi [Monthly Notices of the Royal Astronomical Society] {10.1093/mnras/stz3305}, 491, 4247–4253

\bibitem[\protect\citeauthoryear{Hennawi, Dalal, Bode  \& Ostriker}{Hennawi et~al.}{2007}]{Hennawi2007}
Hennawi J.~F.,  Dalal N.,  Bode P.,   Ostriker J.~P.,  2007, \mn@doi [The Astrophysical Journal] {10.1086/497362}, 654, 714

\bibitem[\protect\citeauthoryear{Hezaveh et~al.,}{Hezaveh et~al.}{2016}]{Hezaveh2016}
Hezaveh Y.~D.,  et~al., 2016, \mn@doi [The Astrophysical Journal] {10.3847/0004-637x/823/1/37}, 823, 37

\bibitem[\protect\citeauthoryear{Hoffmann et~al.,}{Hoffmann et~al.}{2014}]{Hoffmann_2014}
Hoffmann K.,  et~al., 2014, \mn@doi [Monthly Notices of the Royal Astronomical Society] {10.1093/mnras/stu933}, 442, 1197–1210

\bibitem[\protect\citeauthoryear{{Holmberg}}{{Holmberg}}{1969}]{holmberg1969}
{Holmberg} E.,  1969, Arkiv for Astronomi, \href {https://ui.adsabs.harvard.edu/abs/1969ArA.....5..305H} {5, 305}

\bibitem[\protect\citeauthoryear{Hopkins, Bahcall  \& Bode}{Hopkins et~al.}{2005}]{Hopkins2005}
Hopkins P.~F.,  Bahcall N.~A.,   Bode P.,  2005, \mn@doi [The Astrophysical Journal] {10.1086/425993}, 618, 1

\bibitem[\protect\citeauthoryear{{Hui}}{{Hui}}{2021}]{hui2021}
{Hui} L.,  2021, \mn@doi [\araa] {10.1146/annurev-astro-120920-010024}, \href {https://ui.adsabs.harvard.edu/abs/2021ARA&A..59..247H} {59, 247}

\bibitem[\protect\citeauthoryear{Hunter}{Hunter}{2007}]{matplotlib}
Hunter J.~D.,  2007, \mn@doi [Computing in Science Engineering] {10.1109/MCSE.2007.55}, 9, 90

\bibitem[\protect\citeauthoryear{Ibata et~al.,}{Ibata et~al.}{2013}]{Ibata_2013}
Ibata R.~A.,  et~al., 2013, \mn@doi [Nature] {10.1038/nature11717}, 493, 62–65

\bibitem[\protect\citeauthoryear{Ivezic et~al.,}{Ivezic et~al.}{2019}]{Ivezi_2019}
Ivezic Z.,  et~al., 2019, \mn@doi [The Astrophysical Journal] {10.3847/1538-4357/ab042c}, 873, 111

\bibitem[\protect\citeauthoryear{{Jiang} \& {van den Bosch}}{{Jiang} \& {van den Bosch}}{2016}]{jiang2016}
{Jiang} F.,  {van den Bosch} F.~C.,  2016, \mn@doi [\mnras] {10.1093/mnras/stw439}, \href {https://ui.adsabs.harvard.edu/#abs/2016MNRAS.458.2848J} {458, 2848}

\bibitem[\protect\citeauthoryear{Jing}{Jing}{2002}]{Jing_2002}
Jing Y.~P.,  2002, \mn@doi [Monthly Notices of the Royal Astronomical Society] {10.1046/j.1365-8711.2002.05899.x}, 335, L89–L93

\bibitem[\protect\citeauthoryear{Jing \& Suto}{Jing \& Suto}{2002}]{Jing2002}
Jing Y.~P.,  Suto Y.,  2002, \mn@doi [The Astrophysical Journal] {10.1086/341065}, 574, 538

\bibitem[\protect\citeauthoryear{Jones, Oliphant, Peterson  et~al.}{Jones et~al.}{2001}]{scipy}
Jones E.,  Oliphant T.,  Peterson P.,   et~al., 2001, {SciPy}: Open source scientific tools for {Python}, \url {http://www.scipy.org/}

\bibitem[\protect\citeauthoryear{Kang \& Wang}{Kang \& Wang}{2015}]{Kang_2015}
Kang X.,  Wang P.,  2015, \mn@doi [The Astrophysical Journal] {10.1088/0004-637X/813/1/6}, 813, 6

\bibitem[\protect\citeauthoryear{Karp, Lange  \& Wechsler}{Karp et~al.}{2023}]{Karp_2023}
Karp J. S.~M.,  Lange J.~U.,   Wechsler R.~H.,  2023, \mn@doi [The Astrophysical Journal Letters] {10.3847/2041-8213/acd3e9}, 949, L13

\bibitem[\protect\citeauthoryear{{Kauffmann}, {White}  \& {Guiderdoni}}{{Kauffmann} et~al.}{1993}]{kauffmann1993}
{Kauffmann} G.,  {White} S.~D.~M.,   {Guiderdoni} B.,  1993, \mn@doi [\mnras] {10.1093/mnras/264.1.201}, \href {https://ui.adsabs.harvard.edu/#abs/1993MNRAS.264..201K} {264, 201}

\bibitem[\protect\citeauthoryear{Kaufmann, Mayer, Wadsley, Stadel  \& Moore}{Kaufmann et~al.}{2007}]{kaufmann2007}
Kaufmann T.,  Mayer L.,  Wadsley J.,  Stadel J.,   Moore B.,  2007, \mn@doi [Monthly Notices of the Royal Astronomical Society] {10.1111/j.1365-2966.2006.11314.x}, 375, 53

\bibitem[\protect\citeauthoryear{Keeton}{Keeton}{2003}]{Keeton_2003}
Keeton C.~R.,  2003, \mn@doi [The Astrophysical Journal] {10.1086/345717}, 584, 664–674

\bibitem[\protect\citeauthoryear{Kiessling et~al.,}{Kiessling et~al.}{2015}]{Kiessling_2015}
Kiessling A.,  et~al., 2015, \mn@doi [Space Science Reviews] {10.1007/s11214-015-0203-6}, 193, 67–136

\bibitem[\protect\citeauthoryear{{Klypin}, {Kravtsov}, {Valenzuela}  \& {Prada}}{{Klypin} et~al.}{1999}]{klypin1999}
{Klypin} A.,  {Kravtsov} A.~V.,  {Valenzuela} O.,   {Prada} F.,  1999, \mn@doi [\apj] {10.1086/307643}, \href {https://ui.adsabs.harvard.edu/#abs/1999ApJ...522...82K} {522, 82}

\bibitem[\protect\citeauthoryear{Knebe, Draganova, Power, Yepes, Hoffman, Gottlöber  \& Gibson}{Knebe et~al.}{2008}]{Knebe_2008}
Knebe A.,  Draganova N.,  Power C.,  Yepes G.,  Hoffman Y.,  Gottlöber S.,   Gibson B.~K.,  2008, \mn@doi [Monthly Notices of the Royal Astronomical Society: Letters] {10.1111/j.1745-3933.2008.00459.x}, 386, L52–L56

\bibitem[\protect\citeauthoryear{{Kroupa}, {Theis}  \& {Boily}}{{Kroupa} et~al.}{2005}]{kroupa_2005}
{Kroupa} P.,  {Theis} C.,   {Boily} C.~M.,  2005, \mn@doi [\aap] {10.1051/0004-6361:20041122}, \href {https://ui.adsabs.harvard.edu/abs/2005A&A...431..517K} {431, 517}

\bibitem[\protect\citeauthoryear{{Kunkel} \& {Demers}}{{Kunkel} \& {Demers}}{1976}]{kunkel_1976}
{Kunkel} W.~E.,  {Demers} S.,  1976, in The Galaxy and the Local Group. p.~241

\bibitem[\protect\citeauthoryear{Lacey \& Cole}{Lacey \& Cole}{1994}]{Lacey1994}
Lacey C.,  Cole S.,  1994, \mn@doi [Monthly Notices of the Royal Astronomical Society] {10.1093/mnras/271.3.676}, 271, 676

\bibitem[\protect\citeauthoryear{Li, Jing, Faltenbacher  \& Wang}{Li et~al.}{2013}]{Li_2013}
Li C.,  Jing Y.~P.,  Faltenbacher A.,   Wang J.,  2013, \mn@doi [The Astrophysical Journal] {10.1088/2041-8205/770/1/l12}, 770, L12

\bibitem[\protect\citeauthoryear{Libeskind, Cole, Frenk, Okamoto  \& Jenkins}{Libeskind et~al.}{2007}]{Libeskind_2007}
Libeskind N.~I.,  Cole S.,  Frenk C.~S.,  Okamoto T.,   Jenkins A.,  2007, \mn@doi [Monthly Notices of the Royal Astronomical Society] {10.1111/j.1365-2966.2006.11205.x}, 374, 16–28

\bibitem[\protect\citeauthoryear{Libeskind, Knebe, Hoffman, Gottlöber, Yepes  \& Steinmetz}{Libeskind et~al.}{2010}]{Libeskind2010}
Libeskind N.~I.,  Knebe A.,  Hoffman Y.,  Gottlöber S.,  Yepes G.,   Steinmetz M.,  2010, \mn@doi [Monthly Notices of the Royal Astronomical Society] {10.1111/j.1365-2966.2010.17786.x}, 411, 1525–1535

\bibitem[\protect\citeauthoryear{{Libeskind}, {Knebe}, {Hoffman}, {Gottl{\"o}ber}, {Yepes}  \& {Steinmetz}}{{Libeskind} et~al.}{2011}]{libeskind_2011}
{Libeskind} N.~I.,  {Knebe} A.,  {Hoffman} Y.,  {Gottl{\"o}ber} S.,  {Yepes} G.,   {Steinmetz} M.,  2011, \mn@doi [\mnras] {10.1111/j.1365-2966.2010.17786.x}, \href {https://ui.adsabs.harvard.edu/abs/2011MNRAS.411.1525L} {411, 1525}

\bibitem[\protect\citeauthoryear{Libeskind, Tempel, Hoffman, Tully  \& Courtois}{Libeskind et~al.}{2015}]{Libeskind_2015}
Libeskind N.~I.,  Tempel E.,  Hoffman Y.,  Tully R.~B.,   Courtois H.,  2015, \mn@doi [Monthly Notices of the Royal Astronomical Society: Letters] {10.1093/mnrasl/slv099}, 453, L108–L112

\bibitem[\protect\citeauthoryear{{Libeskind}, {Guo}, {Tempel}  \& {Ibata}}{{Libeskind} et~al.}{2016}]{libeskind_2016}
{Libeskind} N.~I.,  {Guo} Q.,  {Tempel} E.,   {Ibata} R.,  2016, \mn@doi [\apj] {10.3847/0004-637X/830/2/121}, \href {https://ui.adsabs.harvard.edu/abs/2016ApJ...830..121L} {830, 121}

\bibitem[\protect\citeauthoryear{{Lovell}, {Frenk}, {Eke}, {Jenkins}, {Gao}  \& {Theuns}}{{Lovell} et~al.}{2014}]{lovell+2014}
{Lovell} M.~R.,  {Frenk} C.~S.,  {Eke} V.~R.,  {Jenkins} A.,  {Gao} L.,   {Theuns} T.,  2014, \mn@doi [\mnras] {10.1093/mnras/stt2431}, \href {https://ui.adsabs.harvard.edu/abs/2014MNRAS.439..300L} {439, 300}

\bibitem[\protect\citeauthoryear{{Lynden-Bell}}{{Lynden-Bell}}{1976}]{lynden-bell_1976}
{Lynden-Bell} D.,  1976, \mn@doi [\mnras] {10.1093/mnras/174.3.695}, \href {https://ui.adsabs.harvard.edu/abs/1976MNRAS.174..695L} {174, 695}

\bibitem[\protect\citeauthoryear{{Majewski}}{{Majewski}}{1994}]{majewski_1994}
{Majewski} S.~R.,  1994, \mn@doi [\apjl] {10.1086/187462}, \href {https://ui.adsabs.harvard.edu/abs/1994ApJ...431L..17M} {431, L17}

\bibitem[\protect\citeauthoryear{Mao \& Schneider}{Mao \& Schneider}{1998}]{Mao_1998}
Mao S.,  Schneider P.,  1998, \mn@doi [Monthly Notices of the Royal Astronomical Society] {10.1046/j.1365-8711.1998.01319.x}, 295, 587–594

\bibitem[\protect\citeauthoryear{Mao, Williamson  \& Wechsler}{Mao et~al.}{2015}]{Mao2015}
Mao Y.-Y.,  Williamson M.,   Wechsler R.~H.,  2015, \mn@doi [The Astrophysical Journal] {10.1088/0004-637x/810/1/21}, 810, 21

\bibitem[\protect\citeauthoryear{{Mateo}}{{Mateo}}{1998}]{mateo_1998}
{Mateo} M.~L.,  1998, \mn@doi [\araa] {10.1146/annurev.astro.36.1.435}, \href {https://ui.adsabs.harvard.edu/abs/1998ARA&A..36..435M} {36, 435}

\bibitem[\protect\citeauthoryear{McKinney}{McKinney}{2010}]{pandas}
McKinney W.,  2010, in van~der Walt S.,  Millman J.,  eds, Proceedings of the 9th Python in Science Conference. pp 51 -- 56

\bibitem[\protect\citeauthoryear{Mckean et~al.,}{Mckean et~al.}{2015}]{mckean2015}
Mckean J.,  et~al., 2015, \mn@doi [PoS] {10.22323/1.215.0084}, AASKA14, 084

\bibitem[\protect\citeauthoryear{Meneghetti, Argazzi, Pace, Moscardini, Dolag, Bartelmann, Li  \& Oguri}{Meneghetti et~al.}{2006}]{Meneghetti2006}
Meneghetti M.,  Argazzi R.,  Pace F.,  Moscardini L.,  Dolag K.,  Bartelmann M.,  Li G.,   Oguri M.,  2006, \mn@doi [Astronomy {\&} Astrophysics] {10.1051/0004-6361:20065722}, 461, 25

\bibitem[\protect\citeauthoryear{Metcalf \& Madau}{Metcalf \& Madau}{2001}]{Metcalf_2001}
Metcalf R.~B.,  Madau P.,  2001, \mn@doi [The Astrophysical Journal] {10.1086/323695}, 563, 9–20

\bibitem[\protect\citeauthoryear{Metcalf \& Zhao}{Metcalf \& Zhao}{2002}]{Metcalf_2002}
Metcalf R.~B.,  Zhao H.,  2002, \mn@doi [The Astrophysical Journal] {10.1086/339798}, 567, L5–L8

\bibitem[\protect\citeauthoryear{Metcalf, Moustakas, Bunker  \& Parry}{Metcalf et~al.}{2004}]{Metcalf_2004}
Metcalf R.~B.,  Moustakas L.~A.,  Bunker A.~J.,   Parry I.~R.,  2004, \mn@doi [The Astrophysical Journal] {10.1086/383243}, 607, 43–59

\bibitem[\protect\citeauthoryear{{Metz}, {Kroupa}, {Theis}, {Hensler}  \& {Jerjen}}{{Metz} et~al.}{2009}]{metz_2009}
{Metz} M.,  {Kroupa} P.,  {Theis} C.,  {Hensler} G.,   {Jerjen} H.,  2009, \mn@doi [\apj] {10.1088/0004-637X/697/1/269}, \href {https://ui.adsabs.harvard.edu/abs/2009ApJ...697..269M} {697, 269}

\bibitem[\protect\citeauthoryear{Mezini, Fielder, Zentner, Mao, Wang  \& Wu}{Mezini et~al.}{2023}]{Mezini_2023}
Mezini L.,  Fielder C.~E.,  Zentner A.~R.,  Mao Y.-Y.,  Wang K.,   Wu H.-Y.,  2023, \mn@doi [Monthly Notices of the Royal Astronomical Society] {10.1093/mnras/stad2929}, 526, 4157–4172

\bibitem[\protect\citeauthoryear{Minor, Gad-Nasr, Kaplinghat  \& Vegetti}{Minor et~al.}{2021}]{Minor2021}
Minor Q.,  Gad-Nasr S.,  Kaplinghat M.,   Vegetti S.,  2021, \mn@doi [Monthly Notices of the Royal Astronomical Society] {10.1093/mnras/stab2247}, 507, 1662

\bibitem[\protect\citeauthoryear{{Moore}, {Ghigna}, {Governato}, {Lake}, {Quinn}, {Stadel}  \& {Tozzi}}{{Moore} et~al.}{1999}]{moore1999}
{Moore} B.,  {Ghigna} S.,  {Governato} F.,  {Lake} G.,  {Quinn} T.,  {Stadel} J.,   {Tozzi} P.,  1999, \mn@doi [\apj] {10.1086/312287}, \href {https://ui.adsabs.harvard.edu/#abs/1999ApJ...524L..19M} {524, L19}

\bibitem[\protect\citeauthoryear{Morinaga \& Ishiyama}{Morinaga \& Ishiyama}{2020}]{Morinaga_2020}
Morinaga Y.,  Ishiyama T.,  2020, \mn@doi [Monthly Notices of the Royal Astronomical Society] {10.1093/mnras/staa1180}, 495, 502–509

\bibitem[\protect\citeauthoryear{Nadler et~al.,}{Nadler et~al.}{2023}]{Nadler_2023}
Nadler E.~O.,  et~al., 2023, \mn@doi [The Astrophysical Journal] {10.3847/1538-4357/acb68c}, 945, 159

\bibitem[\protect\citeauthoryear{{Navarro} \& {Benz}}{{Navarro} \& {Benz}}{1991}]{navarro1991}
{Navarro} J.~F.,  {Benz} W.,  1991, \mn@doi [\apj] {10.1086/170590}, \href {https://ui.adsabs.harvard.edu/abs/1991ApJ...380..320N} {380, 320}

\bibitem[\protect\citeauthoryear{{Navarro} \& {Steinmetz}}{{Navarro} \& {Steinmetz}}{1997}]{navarro1997}
{Navarro} J.~F.,  {Steinmetz} M.,  1997, \mn@doi [\apj] {10.1086/303763}, \href {https://ui.adsabs.harvard.edu/abs/1997ApJ...478...13N} {478, 13}

\bibitem[\protect\citeauthoryear{{Navarro} \& {White}}{{Navarro} \& {White}}{1993}]{navarro1993}
{Navarro} J.~F.,  {White} S.~D.~M.,  1993, \mn@doi [\mnras] {10.1093/mnras/265.2.271}, \href {https://ui.adsabs.harvard.edu/abs/1993MNRAS.265..271N} {265, 271}

\bibitem[\protect\citeauthoryear{Nierenberg et~al.,}{Nierenberg et~al.}{2024}]{keeley+2024}
Nierenberg A.~M.,  et~al., 2024, \mn@doi [Monthly Notices of the Royal Astronomical Society] {10.1093/mnras/stae499}, 530, 2960

\bibitem[\protect\citeauthoryear{Oguri \& Marshall}{Oguri \& Marshall}{2010}]{Oguri_2010}
Oguri M.,  Marshall P.~J.,  2010, \mn@doi [Monthly Notices of the Royal Astronomical Society] {10.1111/j.1365-2966.2010.16639.x}, pp no--no

\bibitem[\protect\citeauthoryear{Oguri et~al.,}{Oguri et~al.}{2012}]{Oguri2012}
Oguri M.,  et~al., 2012, \mn@doi [Monthly Notices of the Royal Astronomical Society] {10.1093/mnras/sts351}, 429, 482

\bibitem[\protect\citeauthoryear{{Okabe} et~al.,}{{Okabe} et~al.}{2020}]{okabe+2020}
{Okabe} T.,  et~al., 2020, \mn@doi [\mnras] {10.1093/mnras/staa1479}, \href {https://ui.adsabs.harvard.edu/abs/2020MNRAS.496.2591O} {496, 2591}

\bibitem[\protect\citeauthoryear{Pawlowski}{Pawlowski}{2018}]{Pawlowski2018}
Pawlowski M.~S.,  2018, \mn@doi [Modern Physics Letters A] {10.1142/s0217732318300045}, 33, 1830004

\bibitem[\protect\citeauthoryear{Pawlowski, Pflamm-Altenburg  \& Kroupa}{Pawlowski et~al.}{2012}]{Pawlowski_2012}
Pawlowski M.~S.,  Pflamm-Altenburg J.,   Kroupa P.,  2012, \mn@doi [Monthly Notices of the Royal Astronomical Society] {10.1111/j.1365-2966.2012.20937.x}, 423, 1109–1126

\bibitem[\protect\citeauthoryear{{Paz}, {Stasyszyn}  \& {Padilla}}{{Paz} et~al.}{2008}]{Paz2008}
{Paz} D.~J.,  {Stasyszyn} F.,   {Padilla} N.~D.,  2008, \mn@doi [\mnras] {10.1111/j.1365-2966.2008.13655.x}, \href {https://ui.adsabs.harvard.edu/abs/2008MNRAS.389.1127P} {389, 1127}

\bibitem[\protect\citeauthoryear{{Peebles}}{{Peebles}}{1969}]{peebles1969}
{Peebles} P.~J.~E.,  1969, \mn@doi [\apj] {10.1086/149876}, \href {https://ui.adsabs.harvard.edu/#abs/1969ApJ...155..393P} {155, 393}

\bibitem[\protect\citeauthoryear{P\'erez \& Granger}{P\'erez \& Granger}{2007}]{ipython}
P\'erez F.,  Granger B.~E.,  2007, \mn@doi [Computing in Science Engineering] {10.1109/MCSE.2007.53}, 9, 21

\bibitem[\protect\citeauthoryear{{Peter} \& {Benson}}{{Peter} \& {Benson}}{2010}]{peter_benson2010}
{Peter} A. H.~G.,  {Benson} A.~J.,  2010, \mn@doi [\prd] {10.1103/PhysRevD.82.123521}, \href {https://ui.adsabs.harvard.edu/abs/2010PhRvD..82l3521P} {82, 123521}

\bibitem[\protect\citeauthoryear{Plionis \& Basilakos}{Plionis \& Basilakos}{2002}]{Plionis_2002}
Plionis M.,  Basilakos S.,  2002, \mn@doi [Monthly Notices of the Royal Astronomical Society] {10.1046/j.1365-8711.2002.05177.x}, 329, L47–L51

\bibitem[\protect\citeauthoryear{{Porciani}, {Dekel}  \& {Hoffman}}{{Porciani} et~al.}{2002a}]{ttt1}
{Porciani} C.,  {Dekel} A.,   {Hoffman} Y.,  2002a, \mn@doi [\mnras] {10.1046/j.1365-8711.2002.05305.x}, \href {https://ui.adsabs.harvard.edu/abs/2002MNRAS.332..325P} {332, 325}

\bibitem[\protect\citeauthoryear{{Porciani}, {Dekel}  \& {Hoffman}}{{Porciani} et~al.}{2002b}]{ttt2}
{Porciani} C.,  {Dekel} A.,   {Hoffman} Y.,  2002b, \mn@doi [\mnras] {10.1046/j.1365-8711.2002.05306.x}, \href {https://ui.adsabs.harvard.edu/abs/2002MNRAS.332..339P} {332, 339}

\bibitem[\protect\citeauthoryear{{Robison} et~al.,}{{Robison} et~al.}{2023}]{robison+2023}
{Robison} B.,  et~al., 2023, \mn@doi [\mnras] {10.1093/mnras/stad1519}, \href {https://ui.adsabs.harvard.edu/abs/2023MNRAS.523.1614R} {523, 1614}

\bibitem[\protect\citeauthoryear{Rodriguez, Merchán  \& Artale}{Rodriguez et~al.}{2022}]{Rodriguez_2022}
Rodriguez F.,  Merchán M.,   Artale M.~C.,  2022, \mn@doi [Monthly Notices of the Royal Astronomical Society] {10.1093/mnras/stac1428}, 514, 1077–1087

\bibitem[\protect\citeauthoryear{Rodriguez, Merchán  \& Artale}{Rodriguez et~al.}{2024}]{Rodriguez_2024}
Rodriguez F.,  Merchán M.,   Artale M.~C.,  2024, Evolution of central galaxy alignments in simulations (\mn@eprint {arXiv} {2405.02398})

\bibitem[\protect\citeauthoryear{Sales \& Navarro}{Sales \& Navarro}{2023}]{Sales2023}
Sales L.~V.,  Navarro J.~F.,  2023, \mn@doi [Nature Astronomy] {10.1038/s41550-023-01924-y}, 7, 376

\bibitem[\protect\citeauthoryear{Samuel, Wetzel, Chapman, Tollerud, Hopkins, Boylan-Kolchin, Bailin  \& Faucher-Gigu{\`{e}}re}{Samuel et~al.}{2021}]{Samuel2021}
Samuel J.,  Wetzel A.,  Chapman S.,  Tollerud E.,  Hopkins P.~F.,  Boylan-Kolchin M.,  Bailin J.,   Faucher-Gigu{\`{e}}re C.-A.,  2021, \mn@doi [Monthly Notices of the Royal Astronomical Society] {10.1093/mnras/stab955}, 504, 1379

\bibitem[\protect\citeauthoryear{{Samuels} \& {Brainerd}}{{Samuels} \& {Brainerd}}{2023}]{samuels_2023}
{Samuels} A.,  {Brainerd} T.~G.,  2023, \mn@doi [\apj] {10.3847/1538-4357/acc069}, \href {https://ui.adsabs.harvard.edu/abs/2023ApJ...947...56S} {947, 56}

\bibitem[\protect\citeauthoryear{Santos-Santos et~al.,}{Santos-Santos et~al.}{2020}]{SantosSantos2020}
Santos-Santos I.,  et~al., 2020, \mn@doi [The Astrophysical Journal] {10.3847/1538-4357/ab7f29}, 897, 71

\bibitem[\protect\citeauthoryear{{Schneider}, {Frenk}  \& {Cole}}{{Schneider} et~al.}{2012a}]{schneider2012}
{Schneider} M.~D.,  {Frenk} C.~S.,   {Cole} S.,  2012a, \mn@doi [Journal of Cosmology and Astro-Particle Physics] {10.1088/1475-7516/2012/05/030}, \href {https://ui.adsabs.harvard.edu/#abs/2012JCAP...05..030S} {2012, 030}

\bibitem[\protect\citeauthoryear{Schneider, Frenk  \& Cole}{Schneider et~al.}{2012b}]{Schneider_2012}
Schneider M.~D.,  Frenk C.~S.,   Cole S.,  2012b, \mn@doi [Journal of Cosmology and Astroparticle Physics] {10.1088/1475-7516/2012/05/030}, 2012, 030

\bibitem[\protect\citeauthoryear{{Shao}, {Cautun}, {Frenk}, {Grand}, {G{\'o}mez}, {Marinacci}  \& {Simpson}}{{Shao} et~al.}{2018}]{shao_2018}
{Shao} S.,  {Cautun} M.,  {Frenk} C.~S.,  {Grand} R. J.~J.,  {G{\'o}mez} F.~A.,  {Marinacci} F.,   {Simpson} C.~M.,  2018, \mn@doi [\mnras] {10.1093/mnras/sty343}, \href {https://ui.adsabs.harvard.edu/abs/2018MNRAS.476.1796S} {476, 1796}

\bibitem[\protect\citeauthoryear{Shaw, Weller, Ostriker  \& Bode}{Shaw et~al.}{2006}]{Shaw_2006}
Shaw L.~D.,  Weller J.,  Ostriker J.~P.,   Bode P.,  2006, \mn@doi [The Astrophysical Journal] {10.1086/505016}, 646, 815–833

\bibitem[\protect\citeauthoryear{{Shi}, {Wang}  \& {Mo}}{{Shi} et~al.}{2015}]{shi2015}
{Shi} J.,  {Wang} H.,   {Mo} H.~J.,  2015, \mn@doi [\apj] {10.1088/0004-637X/807/1/37}, \href {https://ui.adsabs.harvard.edu/abs/2015ApJ...807...37S} {807, 37}

\bibitem[\protect\citeauthoryear{Somerville, Hopkins, Cox, Robertson  \& Hernquist}{Somerville et~al.}{2008}]{Somerville2008}
Somerville R.~S.,  Hopkins P.~F.,  Cox T.~J.,  Robertson B.~E.,   Hernquist L.,  2008, \mn@doi [Monthly Notices of the Royal Astronomical Society] {10.1111/j.1365-2966.2008.13805.x}, 391, 481

\bibitem[\protect\citeauthoryear{{Springel}, {White}  \& {Hernquist}}{{Springel} et~al.}{2004}]{springel2004}
{Springel} V.,  {White} S.~D.~M.,   {Hernquist} L.,  2004, {The shapes of simulated dark matter halos}

\bibitem[\protect\citeauthoryear{Taylor \& Babul}{Taylor \& Babul}{2002}]{taylor2002}
Taylor J.~E.,  Babul A.,  2002, A Semi-analytic Model of Halo Dynamics (\mn@eprint {arXiv} {astro-ph/0201370})

\bibitem[\protect\citeauthoryear{{Taylor} \& {Babul}}{{Taylor} \& {Babul}}{2004}]{taylor_babul2004}
{Taylor} J.~E.,  {Babul} A.,  2004, \mn@doi [\mnras] {10.1111/j.1365-2966.2004.07395.x}, \href {https://ui.adsabs.harvard.edu/abs/2004MNRAS.348..811T} {348, 811}

\bibitem[\protect\citeauthoryear{{Thomas} et~al.,}{{Thomas} et~al.}{1998}]{Thomas_1998}
{Thomas} P.~A.,  et~al., 1998, \mn@doi [\mnras] {10.1046/j.1365-8711.1998.01491.x}, \href {https://ui.adsabs.harvard.edu/abs/1998MNRAS.296.1061T} {296, 1061}

\bibitem[\protect\citeauthoryear{{Tormen}}{{Tormen}}{1997}]{tormen1997}
{Tormen} G.,  1997, \mn@doi [\mnras] {10.1093/mnras/290.3.411}, \href {https://ui.adsabs.harvard.edu/abs/1997MNRAS.290..411T} {290, 411}

\bibitem[\protect\citeauthoryear{Treu \& Marshall}{Treu \& Marshall}{2016}]{Treu_2016}
Treu T.,  Marshall P.~J.,  2016, \mn@doi [The Astronomy and Astrophysics Review] {10.1007/s00159-016-0096-8}, 24

\bibitem[\protect\citeauthoryear{Treu, Suyu  \& Marshall}{Treu et~al.}{2022}]{Treu_2022}
Treu T.,  Suyu S.~H.,   Marshall P.~J.,  2022, \mn@doi [The Astronomy and Astrophysics Review] {10.1007/s00159-022-00145-y}, 30

\bibitem[\protect\citeauthoryear{{Tulin} \& {Yu}}{{Tulin} \& {Yu}}{2018}]{tulin+yu2018}
{Tulin} S.,  {Yu} H.-B.,  2018, \mn@doi [\physrep] {10.1016/j.physrep.2017.11.004}, \href {https://ui.adsabs.harvard.edu/abs/2018PhR...730....1T} {730, 1}

\bibitem[\protect\citeauthoryear{Vegetti, Koopmans, Bolton, Treu  \& Gavazzi}{Vegetti et~al.}{2010}]{Vegetti2010}
Vegetti S.,  Koopmans L. V.~E.,  Bolton A.,  Treu T.,   Gavazzi R.,  2010, \mn@doi [Monthly Notices of the Royal Astronomical Society] {10.1111/j.1365-2966.2010.16865.x}, 408, 1969

\bibitem[\protect\citeauthoryear{{Velliscig} et~al.,}{{Velliscig} et~al.}{2015}]{velliscig+2015}
{Velliscig} M.,  et~al., 2015, \mn@doi [\mnras] {10.1093/mnras/stv1690}, \href {https://ui.adsabs.harvard.edu/abs/2015MNRAS.453..721V} {453, 721}

\bibitem[\protect\citeauthoryear{Vera-Ciro, Sales, Helmi, Frenk, Navarro, Springel, Vogelsberger  \& White}{Vera-Ciro et~al.}{2011}]{Vera_Ciro_2011}
Vera-Ciro C.~A.,  Sales L.~V.,  Helmi A.,  Frenk C.~S.,  Navarro J.~F.,  Springel V.,  Vogelsberger M.,   White S. D.~M.,  2011, \mn@doi [Monthly Notices of the Royal Astronomical Society] {10.1111/j.1365-2966.2011.19134.x}, 416, 1377

\bibitem[\protect\citeauthoryear{Wang, Jing, Mao  \& Kang}{Wang et~al.}{2005}]{Wang_2005}
Wang H.~Y.,  Jing Y.~P.,  Mao S.,   Kang X.,  2005, \mn@doi [Monthly Notices of the Royal Astronomical Society] {10.1111/j.1365-2966.2005.09543.x}, 364, 424–432

\bibitem[\protect\citeauthoryear{{Wang}, {Croft}, {Peter}, {Zentner}  \& {Purcell}}{{Wang} et~al.}{2013}]{wang+2013}
{Wang} M.-Y.,  {Croft} R. A.~C.,  {Peter} A. H.~G.,  {Zentner} A.~R.,   {Purcell} C.~W.,  2013, \mn@doi [\prd] {10.1103/PhysRevD.88.123515}, \href {https://ui.adsabs.harvard.edu/abs/2013PhRvD..88l3515W} {88, 123515}

\bibitem[\protect\citeauthoryear{{Wang}, {Peter}, {Strigari}, {Zentner}, {Arant}, {Garrison-Kimmel}  \& {Rocha}}{{Wang} et~al.}{2014}]{wang+2014}
{Wang} M.-Y.,  {Peter} A. H.~G.,  {Strigari} L.~E.,  {Zentner} A.~R.,  {Arant} B.,  {Garrison-Kimmel} S.,   {Rocha} M.,  2014, \mn@doi [\mnras] {10.1093/mnras/stu1747}, \href {https://ui.adsabs.harvard.edu/abs/2014MNRAS.445..614W} {445, 614}

\bibitem[\protect\citeauthoryear{Wang, Luo, Kang, Libeskind, Wang, Zhang, Tempel  \& Guo}{Wang et~al.}{2018}]{Wang_2018}
Wang P.,  Luo Y.,  Kang X.,  Libeskind N.~I.,  Wang L.,  Zhang Y.,  Tempel E.,   Guo Q.,  2018, \mn@doi [The Astrophysical Journal] {10.3847/1538-4357/aabe2b}, 859, 115

\bibitem[\protect\citeauthoryear{{Warren}, {Quinn}, {Salmon}  \& {Zurek}}{{Warren} et~al.}{1992}]{warren_1992}
{Warren} M.~S.,  {Quinn} P.~J.,  {Salmon} J.~K.,   {Zurek} W.~H.,  1992, \mn@doi [\apj] {10.1086/171937}, \href {https://ui.adsabs.harvard.edu/abs/1992ApJ...399..405W} {399, 405}

\bibitem[\protect\citeauthoryear{{Wechsler}, {Bullock}, {Primack}, {Kravtsov}  \& {Dekel}}{{Wechsler} et~al.}{2002}]{wechsler2002}
{Wechsler} R.~H.,  {Bullock} J.~S.,  {Primack} J.~R.,  {Kravtsov} A.~V.,   {Dekel} A.,  2002, \mn@doi [\apj] {10.1086/338765}, \href {https://ui.adsabs.harvard.edu/#abs/2002ApJ...568...52W} {568, 52}

\bibitem[\protect\citeauthoryear{Welker, Devriendt, Dubois, Pichon  \& Peirani}{Welker et~al.}{2014}]{Welker_2014}
Welker C.,  Devriendt J.,  Dubois Y.,  Pichon C.,   Peirani S.,  2014, \mn@doi [Monthly Notices of the Royal Astronomical Society: Letters] {10.1093/mnrasl/slu106}, 445, L46–L50

\bibitem[\protect\citeauthoryear{{White}}{{White}}{1984}]{white1984}
{White} S.~D.~M.,  1984, \mn@doi [\apj] {10.1086/162573}, \href {http://adsabs.harvard.edu/abs/1984ApJ...286...38W} {286, 38}

\bibitem[\protect\citeauthoryear{{White} \& {Rees}}{{White} \& {Rees}}{1978}]{white1978}
{White} S.~D.~M.,  {Rees} M.~J.,  1978, \mn@doi [\mnras] {10.1093/mnras/183.3.341}, \href {https://ui.adsabs.harvard.edu/#abs/1978MNRAS.183..341W} {183, 341}

\bibitem[\protect\citeauthoryear{Willman, Governato, Dalcanton, Reed  \& Quinn}{Willman et~al.}{2004}]{Willman_2004}
Willman B.,  Governato F.,  Dalcanton J.~J.,  Reed D.,   Quinn T.,  2004, \mn@doi [Monthly Notices of the Royal Astronomical Society] {10.1111/j.1365-2966.2004.08095.x}, 353, 639–646

\bibitem[\protect\citeauthoryear{{Wu}, {Hahn}, {Wechsler}, {Mao}  \& {Behroozi}}{{Wu} et~al.}{2013}]{wu2013}
{Wu} H.-Y.,  {Hahn} O.,  {Wechsler} R.~H.,  {Mao} Y.-Y.,   {Behroozi} P.~S.,  2013, \mn@doi [\apj] {10.1088/0004-637X/763/2/70}, \href {https://ui.adsabs.harvard.edu/abs/2013ApJ...763...70W} {763, 70}

\bibitem[\protect\citeauthoryear{Xia, Kang, Wang, Luo, Yang, Jing, Wang  \& Mo}{Xia et~al.}{2017}]{Xia_2017}
Xia Q.,  Kang X.,  Wang P.,  Luo Y.,  Yang X.,  Jing Y.,  Wang H.,   Mo H.,  2017, \mn@doi [The Astrophysical Journal] {10.3847/1538-4357/aa8d17}, 848, 22

\bibitem[\protect\citeauthoryear{Yang, Van Den~Bosch, Mo, Mao, Kang, Weinmann, Guo  \& Jing}{Yang et~al.}{2006}]{Yang_2006}
Yang X.,  Van Den~Bosch F.~C.,  Mo H.~J.,  Mao S.,  Kang X.,  Weinmann S.~M.,  Guo Y.,   Jing Y.~P.,  2006, \mn@doi [Monthly Notices of the Royal Astronomical Society] {10.1111/j.1365-2966.2006.10373.x}, 369, 1293–1302

\bibitem[\protect\citeauthoryear{{Zel'dovich}}{{Zel'dovich}}{1970}]{zeldovich1970}
{Zel'dovich} Y.~B.,  1970, \aap, \href {https://ui.adsabs.harvard.edu/abs/1970A&A.....5...84Z} {5, 84}

\bibitem[\protect\citeauthoryear{Zelko, Nierenberg  \& Treu}{Zelko et~al.}{2024}]{zelko_2023}
Zelko I.~A.,  Nierenberg A.~M.,   Treu T.,  2024, \mn@doi [Monthly Notices of the Royal Astronomical Society] {10.1093/mnras/stae970}, 531, 885

\bibitem[\protect\citeauthoryear{{Zentner} \& {Bullock}}{{Zentner} \& {Bullock}}{2003}]{zentner2003}
{Zentner} A.~R.,  {Bullock} J.~S.,  2003, \mn@doi [\apj] {10.1086/378797}, \href {https://ui.adsabs.harvard.edu/#abs/2003ApJ...598...49Z} {598, 49}

\bibitem[\protect\citeauthoryear{{Zentner}, {Berlind}, {Bullock}, {Kravtsov}  \& {Wechsler}}{{Zentner} et~al.}{2005a}]{ZentnerBerlind2005}
{Zentner} A.~R.,  {Berlind} A.~A.,  {Bullock} J.~S.,  {Kravtsov} A.~V.,   {Wechsler} R.~H.,  2005a, \mn@doi [\apj] {10.1086/428898}, \href {https://ui.adsabs.harvard.edu/abs/2005ApJ...624..505Z} {624, 505}

\bibitem[\protect\citeauthoryear{{Zentner}, {Kravtsov}, {Gnedin}  \& {Klypin}}{{Zentner} et~al.}{2005b}]{zentner2005}
{Zentner} A.~R.,  {Kravtsov} A.~V.,  {Gnedin} O.~Y.,   {Klypin} A.~A.,  2005b, \mn@doi [\apj] {10.1086/431355}, \href {https://ui.adsabs.harvard.edu/#abs/2005ApJ...629..219Z} {629, 219}

\bibitem[\protect\citeauthoryear{{Zhang}, {Yang}, {Faltenbacher}, {Springel}, {Lin}  \& {Wang}}{{Zhang} et~al.}{2009}]{zhang2009}
{Zhang} Y.,  {Yang} X.,  {Faltenbacher} A.,  {Springel} V.,  {Lin} W.,   {Wang} H.,  2009, \mn@doi [\apj] {10.1088/0004-637X/706/1/747}, \href {https://ui.adsabs.harvard.edu/abs/2009ApJ...706..747Z} {706, 747}

\bibitem[\protect\citeauthoryear{van~der Walt, Colbert  \& Varoquaux}{van~der Walt et~al.}{2011}]{numpy}
van~der Walt S.,  Colbert S.~C.,   Varoquaux G.,  2011, \mn@doi [Computing in Science Engineering] {10.1109/MCSE.2011.37}, 13, 22

\bibitem[\protect\citeauthoryear{van den Bosch \& Jiang}{van den Bosch \& Jiang}{2016}]{vandenbosch2014}
van den Bosch F.~C.,  Jiang F.,  2016, \mn@doi [Monthly Notices of the Royal Astronomical Society] {10.1093/mnras/stw440}, 458, 2870

\makeatother
\end{thebibliography}

\end{document}